\documentclass[twocolumn,aps,pra]{revtex4}
\usepackage{epsfig}
\usepackage{amsmath}
\usepackage{array}
\usepackage{amsfonts}
\usepackage{color}
\usepackage{soul}
\usepackage{multirow}
\usepackage{amssymb}

\def\a{\hat{c}}
\def\AAA{\mathbb{A}}
\def\BBB{\bar{B}}

\def\DD{\mathfrak{D}}
\def\DDD{\mathbb{D}}
\def\EE{\bar{\mathcal{E}}}
\def\EEE{\mathbb{E}}

\def\FFF{\mathbb{F}}
\def\GG{\bar{G}}
\def\GGG{\mathbb{G}}
\def\g{\Omega}
\def\HH{\hat{\mathcal{H}}}
\def\HHH{\mathbb{H}}
\def\III{\mathbb{I}}

\def\KKK{\mathbb{K}}
\def\LL{\hat{\hat{\mathcal{L}}}}
\def\LLL{\mathbb{L}}

\def\MM{\mathcal{M}} 
\def\MMM{\mathbb{M}}

\def\QQQ{\mathbb{Q}}

\def\SSS{\bar{S}}
\def\XX{\bar{X}}

\def\bbeta{\bar{\beta}}
\def\hro{\hat{\rho}}
\def\sig{\hat{\sigma}}

\begin{document}

\title{Stability analysis for bad cavity lasers using inhomogeneously broadened spin-1/2 atoms as gain medium}
\author{G.~A.~Kazakov\footnote{E--mail:
kazakov.george@gmail.com}, T.~Schumm}
\affiliation{{\setlength{\baselineskip}{18pt}
{Vienna Center for Quantum Science and Technology (VCQ), Atominstitut, TU Wien, Stadionallee 2, 1020 Vienna, Austria}\\
}}

\begin{abstract}
Bad cavity lasers are experiencing renewed interest in the context of active optical frequency standards, due to their enhanced robustness against fluctuations of the laser cavity. The gain medium would consist of narrow-linewidth atoms, either trapped inside the cavity or intersecting the cavity mode dynamically. A series of effects like the atoms finite velocity distribution, atomic interactions, or interactions of realistic multilevel atoms with auxiliary or stray fields can lead to an inhomogeneous broadening of the atomic gain profile. This causes the emergence of instable regimes of laser operation, characterized by complex temporal patterns of the field amplitude. We study the steady-state solutions and their stability for the metrology-relevant case of a bad cavity laser with spin-1/2 atoms, such as ${\rm ^{171}Yb}$, interacting with an external magnetic field. For the stability analysis, we present a new and efficient method, that can be applied to a broad class of single-mode bad cavity lasers with inhomogeneously broadened multilevel atoms acting as gain medium.
\end{abstract}


\maketitle                

\section{Introduction}
The bad cavity laser is a laser configuration where the linewidth of the cavity mode is spectrally broader than the gain profile of the active medium. The output frequency emitted by such a laser is determined primarily by the properties of the gain medium, it is rather robust to mechanical or thermal fluctuations of the cavity. This opens the possibility to create a highly stable source of radiation, an {\em active optical frequency standard} using narrow-line transition atoms as gain medium. Such standards have been proposed by several authors recently~\cite{Chen05, Meiser09, Yu08, Zhang131, Zhuang14F}. Theoretical estimations~\cite{Meiser09} show that a bad cavity laser using  $10^6$ alkali-earth atoms confined within an optical lattice potential can reach a linewidth down to 1\,mHz. This is more than 1 order of magnitude narrower than what can be realized with the best modern macroscopic resonators ~\cite{Nicholson12,Kessler12,Haefner14}. The development of an active optical frequency standards would be of great relevance for quantum metrology and further applications. To date, such a standard has not been realized mainly due to technical challenges~\cite{Kazakov14},  however a series of proof-of-principle experiments has been performed~\cite{Bohnet12, Bohnet121, Bohnet13, Cox14, Weiner15, Norcia15, Norcia16}.

To realize an ultra-stable bad cavity laser, the active atoms must be confined to the Lamb-Dicke regime to avoid Doppler and recoil shifts. This confinement may be realized with optical lattice potentials formed by counter-propagating laser beams at the so-called ``magic'' wavelength, where the upper and lower lasing states experience the same light shift~\cite{Derevianko11}. These light shifts depend on the polarization of the trapping fields and can be controlled to a certain extent only. Fortunately, for ${^3P_0} \rightarrow {^1S_0}$ transitions in Sr and other alkali-earth atoms, Zn, Cd, Hg, and Yb, this polarization dependence is weak enough, and the relative light shift can be controlled to a high level of precision. Still, active atoms must be continuously repumped to the upper lasing state~\cite{Meiser09, Bohnet14}, and special measures to compensate atom losses from the optical lattice potential must be implemented~\cite{Kazakov13, Kazakov14}. 

A less complex but presumably also less precise optical frequency standard based on continuously pumped active atoms, contained in a thermal vapor cell, placed inside a bad cavity, has been proposed in~\cite{Zhang131,Zhuang14F, XuChen151,XuChen152}. Such a system may be realised as transportable unit for metrology applications outside the physics laboratory.

In both proposed systems and in other possible implementations of bad cavity lasers, {\em inhomogeneous broadening} of the gain profile may occur, for example, through the spatial inhomogeneity of the light shifts caused by the pumping lasers, density- or lattice-induced shifts in trapped atoms, and other possible mechanisms. Additionally, considering a gain medium formed by real multilevel atoms, the lasing states may be split, for example, due to the Zeeman effect.

These broadenings and splittings may considerably alter the properties of the output laser radiation, such as the power and the linewidth. Moreover, they may drastically change the character of lasing. Particularly, the inhomogeneous broadening facilitates transition to the so-called {\em instable regime}~\footnote{Here and below, the term ``(in)stability'' is used as a qualitative characteristic of the temporal behaviour of the laser amplitude and inversion, not for the Allan deviation of the laser frequency.}, where the amplitude of the output laser radiation exhibits strong temporal variations~\cite{Abraham851, Haken}. It can be accompanied by a significant enhancement of phase fluctuations. Phase locking a secondary laser to such an instable source will require special efforts, if possible at all. The development of novel active optical frequency standards must hence include a stability analysis.

This paper is dedicated to the theoretical study of the influence of inhomogeneous broadening on the output power and stability of a single mode bad cavity laser where the gain atoms have split lasing states. In section \ref{sec:method} we present a very general form of the semiclassical equations describing such a system, and introduce an efficient method for the stability analysis. In section~\ref{sec:Yb}, we consider the simplest realistic example of a bad cavity laser with multilevel atoms, namely the optical lattice laser with $\pi$-polarized laser mode, where both lasing states of the active atoms have total angular momentum $F=1/2$. Such a configuration can be realized, for example, with $^{199}$Hg and $^{171}$Yb atoms. We specify our generic semiclassical model for such a system, study the dependence of the attainable steady-state output power, and investigate the stability of these steady-state solutions for various inhomogeneous linewidths and differential Zeeman splittings of the lasing transitions. In section~\ref{sec:outlook} we discuss other possible implementations of bad cavity lasers with simultaneous lasing on different transitions interacting with the same cavity mode, as well as bad cavity lasers with inhomogeneously broadened gain.

\section{General model and method of stability analysis}
\label{sec:method}

In this section we construct a generic form of the semiclassical equations describing the single-mode laser with a gain consisting of multilevel atoms, and present our method for the stability analysis of the steady-state solutions of these equations.

 There are three main approaches to the analysis of laser stability. The first one is the {\em sideband approach}~\cite{Casperson80, Casperson81, Meziane07}, where the Maxwell-Bloch equations, describing the laser, are Fourier-transformed. Instability takes place, if a side mode has a net gain exceeding its losses. The second approach, the {\em linear stability analysis} (LSA), is based on constructing a matrix of Maxwell-Bloch equations linearized near the steady-state solution, and verifying that no eigenvalue of this matrix has a positive real part. It has been shown in~\cite{Mandel85, Abraham85, Zhang85} that the LSA and the sideband approach are formally equivalent. The third approach is based on the direct numerical simulation of the Maxwell-Bloch equations~\cite{Casperson85}.  It allows us to study the temporal behaviour of the laser field amplitude, polarization of the gain medium, and other parameters, but requires extensive computational resources.

The stability analysis can be performed analytically in some particular cases, such as lasers with active two-level atoms with Lorentzian and Gaussian broadening profiles~\cite{Abraham85, Zhang85, Mandel85}, gas laser with active two-level atoms and a saturable absorber~\cite{Salomaa73}, four-level lasers with pump modulation~\cite{Chakmakjian89}, and a few other examples. However, an analytic treatment becomes infeasible for more realistic models of the active medium involving the full atomic level structure, the sideband structure of the optical lattice potential, various inhomogeneous effects, etc. In such cases, the stability analysis has to be performed numerically. A straightforward application of the LSA approach is to partition the gain profile into a finite number of bins, replacing the continuous distribution of atomic frequencies by a discrete one, to linearise the respective set of equations near the steady-state solution, and to calculate the eigenvalues of the matrix of this linearized system numerically. This partitioning should obviously be fine enough to avoid numerical artefacts. The computational cost of the eigenvalue problem (for the desired precision) generally scales {\em cubic} with the number of partitions, which makes the procedure very time-consuming, especially for complex multilevel atoms and significant inhomogeneous broadening. 

A considerable reduction of the computation cost may be attained, if one will focus on the search for the rightmost eigenvalues instead of all eigenvalues. This search may be performed, for example, with the help of the Arnoldi algorithm with Caley transform or Chebyshev iteration~\cite{Meerbergen96}. We should note, however, that these methods should be implemented with care, to avoid too slow convergence and/or missing the rightmost eigenvalue.

In this paper we propose an alternative method, that also does not require solving the complete eigenvalue problem for the linearized system. As it will be shown, its computation cost is {\em linear} in the number of iterations. In the section~\ref{sec:gmod} we introduce the basic assumptions, and derive a very generic form of the semiclassical equations describing the dynamics of the single mode laser. In section~\ref{sec:lsa} we describe the essence of the method. In section~\ref{sec:practLSA} we discuss some details of its practical implementation.

\subsection{Basic assumptions and general form of the semiclassical equations}
\label{sec:gmod}

We consider an ensemble of $N$ pumped (inverted) atoms interacting with a single cavity mode. We suppose that these atoms are confined in space (for example, in an optical lattice potential, or in a solid-state matrix), or the cavity field and pumping fields are running waves, and recoil effects can be neglected. Also we neglect dipole-dipole interaction between the atoms, as well as their collective coupling to the bath modes (the role of these effects on the dynamics of the bad cavity optical lattice laser has been considered in~\cite{Maier14, Kraemer15}). These assumptions, together with a resonance approximation, allow us to eliminate the explicit temporal dependence from the Hamiltonian by transformation into the respective rotating frame. Then one can write the master equation describing the evolution of the system as
\begin{align}
\frac{d\hro}{dt}&=-\frac{i}{\hbar}\left[\HH^0, \hro \right]+ \LL_c[\hro]+\sum_j \LL_{j}[\hro], \label{equ:1}
\end{align}
where the Liouvillian $\LL_j$ describes the relaxation of the $j$th atom,
\begin{align}
\LL_c[\hro]&=-\frac{\kappa}{2} \left[ \a^+ \a\, \hro+\hro\, \a^+\a-2\, \a\, \hro\, \a^+ \right] \label{equ:2}
\end{align}
describes the relaxation of the cavity field, and the Hamiltonian $\HH^0$ may be presented as
\begin{equation}
\HH^0=\HH_c^0+\sum_{j=1}^N \HH^{(j)}_{a,0}+\sum_{j=1}^N \HH_{af}^{(j)}.  \label{equ:3}
\end{equation}
Here the sums are taken over individual atoms, the first term $\HH_c^0=\hbar \omega_c \, \a^+\a$ corresponds to the eigenenergy of the cavity mode, the second term is a sum of single-atom Hamiltonians $\HH^{(j)}_{a,0}$ (which may include interactions with pumping fields, if relevant), and the last term is a sum of Hamiltonians $\HH^{(j)}_{af}$ describing the interaction of $i$th atom with the cavity field:
\begin{equation}
\HH^{(j)}_{af}=\frac{\hbar}{2} \sum_{g,e}\left(\g_{ge}^j \a^+\sig^j_{ge}+\g_{ge}^{j\,*}\sig^j_{eg}\a \right),  \label{eq:4}
\end{equation}
where the sum is taken over sublevels $|g^j\rangle$ and $|e^j\rangle$ of the lower and upper lasing states pertaining to the $j$th atom respectively, $\sig_{xy}^j=|x^j\rangle\langle y^j|$, and $\g^j_{ge}$ is the coupling strength between the $g$th lower and $e$th upper lasing states of the $j$th atom and the cavity field. 

Applying a unitary transformation $\hat{U}=\exp \left[-i\omega t \left(\a^+\a+\sum_{j,g}\sig^j_{gg}\right) \right]$, and introducing the field detuning $\delta=\omega_c-\omega$, we transform the Hamiltonian (\ref{equ:3}) into the form
\begin{equation}
\HH=\hat{U}^+ \HH^0 \hat{U}-i\hbar\, \hat{U}^+\frac{\partial \hat{U}}{\partial t}=\HH_c+\sum_j(\HH^{(j)}_{a}+\HH^{(j)}_{af}), \label{eq:5}
\end{equation}
where
\begin{align}
\HH_c&=\hbar  \delta\,  \a^+\a \label{eq:6}, \\
\HH_{a}^{(j)}&=\HH_{a,0}^{(j)}-\hbar (\omega_c-\delta)\sum_u\sig^j_{uu}.
\label{eq:7}
\end{align}
Note that the transformation (\ref{eq:5}) does not modify the form of the Hamiltonians $\HH^{(j)}_{af}$.

Now we can write the equations of motion for the relevant expectation values of the atomic and the field operators using $\langle \dot{\hat{O}}\rangle={\rm Tr}[\dot{\hat{\rho}}\, \hat{O}]$. In this paper we use the {\em semiclassical approximation}, where the atom-field correlators are factorized, i.e., $\langle \sig_{xy}^j \a \rangle$ is replaced by $\langle \sig_{xy}^j \rangle\langle  \a \rangle$ etc. Using the normalization condition $\sum_{x}\langle \sigma_{xx}^j\rangle=1$, we can represent the set of equations for atomic and field expectation values in matrix form:
\begin{align}
\frac{d\langle \a \rangle}{dt}&=-\left(\frac{\kappa}{2}+i\delta \right)\langle \a \rangle+
\sum_j \GG^{\prime (j)} \cdot \overline{\langle{\sig^j}\rangle} \label{eq:8} \\
\frac{d\left\langle \a^+ \right\rangle}{dt}&=-\left(\frac{\kappa}{2}-i\delta \right)\langle \a^+ \rangle+
\sum_j\GG^{\prime \prime (j)} \cdot \overline{\langle{\sig^j}\rangle} \label{eq:9} \\
\frac{d\overline{\langle \sig^j \rangle}}{dt}&=
\AAA^{(j)} (\delta, \langle \a \rangle, \langle \a^+ \rangle) \cdot \overline{\langle \sig^j \rangle} +\BBB^{(j)} (\langle \a \rangle, \langle \a^+ \rangle) . \label{eq:10}
\end{align}
Here column- and row vectors are indicated by an overline (in particular, $\overline{\langle{\sig^j}\rangle}$ denotes the column vector constructed on expectations $\langle{\sig^j_{xy}}\rangle$ of single-atom operators), matrices are denoted by double-barred letters, the group of equations for $\langle \sig^j_{xy} \rangle$ at specific $j$ is represented as a set of linear differential equations with matrix $\AAA^{(j)}$ and a constant term $\BBB^{(j)}$ appearing due to the normalization condition, row vectors $\GG^{\prime (j)}$ and $\GG^{\prime \prime (j)}$ are defined by
\begin{align}
\GG^{\prime (j)} \cdot \overline{\langle{\sig^j}\rangle}&=
-\frac{i}{2} \sum_{g,e}\g^j_{ge} \langle \sig^j_{ge} \rangle, 
\label{eq:11} \\
\GG^{\prime \prime (j)} \cdot \overline{\langle{\sig^j}\rangle}&=\frac{i}{2} \sum_{g,e}\g^{j\,*}_{ge} \langle \sig^j_{eg} \rangle, \label{eq:12}
\end{align}
and `` $\cdot$ '' denotes an ordinary dot-product. The matricies $\AAA^{(j)}$ and the vectors $\BBB^{(j)}$ depend on $\langle \a \rangle$ and $\langle \a^+ \rangle$ linearly:
\begin{align}
\AAA^{(j)}&=\AAA^{(j)}_0+ \langle \a \rangle \DDD^{\prime (j)} + \langle \a^+ \rangle \DDD^{\prime \prime (j)}, \label{eq:13} \\
\BBB^{(j)}&=\BBB^{(j)}_0+ \langle \a \rangle \bbeta ^{\prime (j)} + \langle \a^+ \rangle \bbeta^{\prime \prime (j)}. \label{eq:14}
\end{align}

Thus, under the assumptions mentioned in the beginning of this section, the semiclassical equations describing the laser dynamics may be presented in the form (\ref{eq:8}) -- (\ref{eq:10}), where $\AAA^{(j)}$ and $\BBB^{(j)}$ have the form (\ref{eq:13}) and (\ref{eq:14}) respectively.

\subsection{Linearization and stability analysis}
\label{sec:lsa}

Suppose we have found a steady-state solution of the equations (\ref{eq:8}) -- (\ref{eq:10}), i.e. the values of $\delta$, $\langle \a \rangle_{cw}=E$, $\langle \a^+ \rangle_{cw}=E^*$, and $\overline{\langle \sig^j \rangle}_{cw}=\SSS^j$ so that, when substituted into the right part of equations (\ref{eq:8}) -- (\ref{eq:10}), we obtain zeros. We now add small perturbations $\varepsilon=\langle \a \rangle-E$, $\varepsilon^*=\langle \a^+ \rangle-E^*$, and $\bar{\chi}^j=\overline{\langle \sig^j \rangle}-\SSS^j$ to the steady-state values of the field and atomic variables. The linearized equations for these perturbations are:

\begin{eqnarray}
\frac{d\varepsilon}{dt}&=&-\left(\frac{\kappa}{2}+i\delta \right)\varepsilon+
\sum_j \GG^{\prime (j)} \cdot \bar{\chi}^j, \label{eq:15}\\
\frac{d\varepsilon^*}{dt}&=&-\left(\frac{\kappa}{2}-i\delta \right)\varepsilon^*+
\sum_j \GG^{\prime \prime (j)} \cdot \bar{\chi}^j, \label{eq:16}\\
\frac{d\bar{\chi}^j}{dt}&=& \AAA^{(j)}\cdot \bar{\chi}^j+(\DDD^{(j)\,\prime} \cdot \SSS^j+\bbeta^{\prime(j)}) \varepsilon \nonumber \\
&&\hspace{11mm}+(\DDD^{j\,\prime \prime} \cdot \SSS^j+\bbeta^{\prime \prime (j)}) \varepsilon^*.  \label{eq:17} 
\end{eqnarray}
It is convenient to introduce the matrices 
\begin{align}
\KKK&= \left(\begin{array}{cc} -\left( \frac{\kappa}{2}+i\delta \right) & 0 \\
0 & -\left( \frac{\kappa}{2}-i\delta \right) \end{array} \right), \label{eq:18}\\
\GGG^{(j)}&=\left( 
\begin{array}{c} \GG^{\prime(j)} \\ \GG^{\prime \prime (j)} \end{array} \right), \label{eq:19} \\
\DDD^{(j)}&=\left(\DDD^{j\,\prime} \cdot \SSS^j+\bbeta^{\prime(j)}\, , \,\DDD^{ \prime \prime (j)} \cdot \SSS^j+\bbeta^{\prime \prime (j)} \right), \label{eq:20}
\end{align}
and the column vector 
\begin{equation}
\EE=\left(\begin{array}{c} \varepsilon \\ \varepsilon^* \end{array} \right). \label{eq:21}
\end{equation}
Then the equations (\ref{eq:15}) -- (\ref{eq:17}) may be written in  matrix form as 
\begin{align}
\frac{d \XX}{dt}&=\LLL \cdot \XX, \label{eq:22}
\end{align}
where the vector $\XX$ and matrix $\LLL$ can be written in block form as
\begin{align}
\XX&=\left[
\begin{array}{c}
\EE \\
\hline  \bar{\chi}^1 \\
\vdots \\ \bar{\chi}^N
\end{array}
 \right], \,
\LLL=\left[ 
\begin{array}{c|ccc}
\KKK & \hphantom{a}\GGG^{(1)} & \hdotsfor{1} & \GGG^{(N)} \\ \hline 
\DDD^{(1)} & \AAA^{(1)} & \dots & 0 \\ 
\vdots & \vdots & \hdotsfor{1} & \vdots  \\
\DDD^{(N)} & 0 & \dots & \AAA^{(N)} \\ 
\end{array}
\right]. \label{eq:23}
\end{align}

To perform the linear stability analysis, it is necessary to  check, whether the matrix $\LLL$ has any eigenvalue with a positive real part, or not. Note that the matrix $\LLL$ has a so-called {\em block arrowhead} structure. Using a well-known theorem about determinants of block matrices \cite{Powell11}
\begin{equation}
\left|\begin{array}{c|c}
\AAA & \mathbb{B} \\ \hline
\mathbb{C} & \DDD
\end{array}\right|
=|\DDD| \, \left|\AAA-\mathbb{B} \cdot \DDD^{-1} \cdot \mathbb{C} \right|, \label{eq:24}
\end{equation}
we represent the characteristic polynomial $|\lambda \III-\LLL|$ of the matrix $\LLL$ as
\begin{equation}
|\lambda \III-\LLL|=\DD(\lambda)\, \left|\lambda \III-\KKK\right| \prod_{j=1}^N |\lambda \III-\AAA^{(j)}|,  \label{eq:25}
\end{equation}
where $\III$ is an identity matrix of the necessary dimension, and
\begin{equation}
\DD(\lambda)=\frac{\displaystyle{\left| \lambda \III-\KKK - \sum_{j=1}^N\GGG^{(j)}\cdot (\lambda \III-\AAA^{(j)})^{-1} \cdot \DDD^{(j)}\right|}}{\left|\lambda \III-\KKK\right|}.
\label{eq:26}
\end{equation}

The function $\DD(\lambda)$ is the cornerstone of our method for the linear stability analysis. First, as one may see from (\ref{eq:25}), $\DD(\lambda)$ is a rational function of $\lambda$, whose numenator is the characteristic polynomial of $\LLL$, and the denumenator is the characteristic polynomial of the matrix $\LLL^\prime$ obtained from $\LLL$ by removal of the non-diagonal blocks. The matrix $\LLL^{\prime}$ describes the dynamics of the damped atoms in a given external field, and the dynamics of the damped field in the medium with a given polarization. The steady-state solutions of the corresponding equations are always stable, therefore all the eigenvalues of the matrix $\LLL^\prime$ (which are the poles of $\DD(\lambda)$) have negative real parts. Both the numenator and denumenator of $\DD(\lambda)$ are polynomials of the same degree in $\lambda$.

{\em To check the laser stability, it is necessary to trace the variation of $\arg(\DD(\lambda))$ over the imaginary axis}. This variation is zero, if all the roots of $\DD(\lambda)$ are located in the left semiplane, or divisible by $2\pi$, if there are some roots in the right semiplane. Note that for non-zero steady-state solution, the point $\lambda=0$ is an eigenvalue of $\LLL$ corresponding to the arbitrary choice of the phase of the laser field. This point should be encircled by a small counterclockwise semicircle, as it is shown in Figure~\ref{fig:f1}, or by some equivalent contour.

\subsection{Practical implementation}
\label{sec:practLSA}

\begin{figure}
\begin{center}
\resizebox{0.45\textwidth}{!}
{\includegraphics{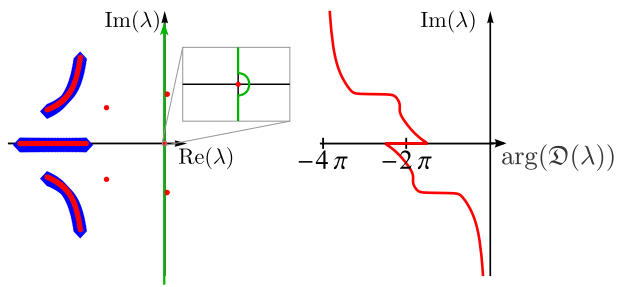}}
\end{center}
\caption{(color online) Illustration of the stability analysis by means of tracing the variation of $\arg(\DD(\lambda))$. Left: $\lambda$ complex plane with roots (red dots) and poles (blue diamonds) of the function $\DD(\lambda)$, and the contour over which to trace the phase (thick green curve). Inset: scaled view of the region around the point $\lambda=0$. Right: Dependence of $\arg(\DD)$ on ${\rm Im}(\lambda)$ along the contour. Here 2 eigenvalues have positive real parts (which indicates instability), and the change of the phase over the contour is equal to $-4\pi$.}
\label{fig:f1}
\end{figure}

Tracing the $\arg{\DD(\lambda)}$ along the contour shown in Figure~\ref{fig:f1} allows us to perform the linear stability analysis without explicit calculation of eigenvalues of the matrix $\LLL$. Usually, the sum over individual atoms should be replaced by an integration over the distributions of the corresponding varying parameters of the atomic ensemble:
\begin{equation}
\begin{split}
&\DD(\lambda)=|\lambda \III - \KKK|^{-1} \left| \lambda \III-\KKK  \vphantom{\frac{dN}{d\Delta_j}d\Delta_j}  \right. \\
&\left. - \int \GGG(\Delta_j) \cdot (\lambda \III - \AAA (\Delta_j))^{-1} \cdot \DDD(\Delta_j) \frac{dN}{d\Delta_j}d\Delta_j\right|, 
\end{split} \label{e:47}
\end{equation}
where $\frac{dN}{d\Delta_j}$ is the distribution of the atoms over the varying parameter $\Delta_j$. Here we introduced $\GGG(\Delta_j)=\GGG^{(j)}$, $\DDD(\Delta_j)=\DDD^{(j)}$, $\AAA(\Delta_j)=\AAA^{(j)}$. For some simple systems and special profiles of inhomogeneous broadening, the integration in (\ref{e:47}) may be performed analytically. In Appendix~\ref{sec:appB} we implement such an analytical treatment to the stability analysis  of a bad cavity optical lattice laser with inhomogeneously broadened, incoherently pumped two-level atoms.

More complex systems, such as multilevel atoms, require numerical integration. It means that the atomic ensemble must be partitioned into $n$ groups, within one group all parameters of the atoms are assumed equal, and the integration is replaced by summing over these partitions:
\begin{equation}
\DD(\lambda)=\frac{\displaystyle{\left| \lambda \III-\KKK - \sum_{k=1}^n N_{k} \GGG^{(k)}\cdot (\lambda \III-\AAA^{(k)})^{-1} \cdot \DDD^{(k)}\right|}}{\left[\left(\frac{\kappa}{2}+\lambda \right)^2+\delta^2\right]},
\label{e:27}
\end{equation}
where $N_k$ is the number of atoms within the $k$th group. The partitioning must be fine enough to avoid unphysical artifacts. It means that the eigenvalues and eigenvectors of the matrices $\AAA^{(j)}$, characterizing individual atoms, as well as the matrices $\GGG^{(j)}$ and $\DDD^{(j)}$, should not differ significantly within one group. For an inhomogeneously broadened atomic ensemble (where the parameter $\Delta_j$ is a detuning of the lasing transition and the individual atoms) that means that the inhomogeneous broadening within single a partition should be smaller than its homogeneous broadening. 

To trace the phase of $\DD(\lambda)$, it is necessary to calculate it in different points of the contour. To reduce the amount of calculations, it is convenient to perform once the eigendecomposition of the matrices $\AAA^{(k)}$, when possible:
\begin{equation}
\AAA^{(k)}= \QQQ^{(k)} \EEE^{(k)} \QQQ^{(k)-1},
\label{e:28}
\end{equation}
where $\EEE^{(k)}$ is a diagonal matrix with eigenvalues of $\AAA^{(k)}$ on the main diagonal, and $\QQQ^{(k)}$ is a square matrix whose columns are the eigenvectors of $\AAA^{(k)}$. Then one can calculate matrices
\begin{equation}
\FFF^{(k)}= \GGG^{(k)} \cdot \QQQ^{(k)}; \quad \HHH^{(k)}= \QQQ^{(k)-1} \cdot \DDD^{(k)}. 
\label{e:29}
\end{equation}
These matrices should be calculated once for every group of atoms, and must be kept in memory. Then 
\begin{equation}
\DD(\lambda)=\frac{\displaystyle{\left| \lambda \III-\KKK - \sum_{k=1}^n N_{k} \FFF^{(k)}\cdot (\lambda \III-\EEE^{(k)})^{-1} \cdot \HHH^{(k)}\right|}}{\left[\left(\frac{\kappa}{2}+\lambda \right)^2+\delta^2\right]}.
\label{e:30}
\end{equation}
The computational cost of standard eigenvalue solvers (for example, reduction to Hessenberg matrix and iterative QR decomposition) scales as the cube of the number of rows of the matrix, therefore the computational cost to evaluate all the matrices $\FFF^{(k)}$, $\EEE^{(k)}$ and $\HHH^{(k)}$ scales as $O(n \times m^3)$, where $n$ is the number of partitions, and $m$ is the number of rows in the matrix $\AAA$. Calculation of $(\lambda \III-\EEE^{(k)})^{-1}$ is trivial and may be easily performed many times. Therefore the  total computational cost our method scales as $O(n \times m^3)+o(m^3)\times n \times n_{g}$, where $n_g$ is a number of nodes discretizing the contour the phase is traced over. In contrast, the computational cost of the standard eigenvalues solver applied directly to the matrix $\LLL$ scales as $O(n^3 \times m^3)$; this difference  becomes especially important for fine-grained partitioning with large $n$.

A few words about numerical tracing of the phase. First, the length of the contour should be chosen, for the sake of confidence, several times larger than the span of the imaginary parts of the eigenvalues of the matrices $\AAA^{(j)}$. Then one needs to create some initial grid on this contour, i.e. to select some set of nodes where $\arg(\DD(\lambda))$ will be calculated. After that, it is necessary to supplement this grid by introducing additional nodes whenever the change of the phase among two adjacent nodes exceeds some level of tolerance, not to miss flips of the phase. For the determination of critical values of some parameters characterising the laser, i.e. the value at which the system looses its stability, it is convenient to adapt the grid by placing more initial nodes into the area where the phase gradient was maximal in the previous step, and less nodes into the other regions of the contour. This can be possible, if the change of parameter per iteration is small enough. Also one can search pure imaginary roots of the function $\DD(\lambda)$, varying both the parameter of interest and ${\rm Im(\lambda)}$.

\section{Optical lattice laser with incoherently pumped spin-1/2 atoms}
\label{sec:Yb}

In this section we consider an optical lattice laser with spin-1/2 alkaline-earth-like atoms, such as ${\rm ^{199}Hg}$, ${\rm ^{111}Cd}$, ${\rm ^{113}Cd}$, and ${\rm ^{171}Yb}$. The latter is a particularly promising candidate for the role of the gain medium in an active optical frequency standard. First, all transitions necessary for cooling and manipulation of ytterbium are in a convenient frequency range, in contrast to mercury and cadmium. Second, the dipole moment of the ${^{1}S_0}\leftrightarrow{^{3}P_0}$ transition in ${\rm ^{171}Yb}$ is much larger than in other alkali-earth-like atoms except the fermionic isotopes of mercury~\cite{Porsev04,Garstang62}. This fact may be considered as a disadvantage for a high-precision {\em passive} optical frequency standard, as a higher dipole moment leads to a broader natural linewidth. However, in {\em active} standards the spectroscopic linewidth is not bounded from below by the natural linewidth. On the contrary, stronger coupling between the atoms and the cavity field allows one to reach the desired linewidth and the output power with a smaller density of atoms, which reduces the collisional shifts. Third, the clock transition in ${\rm ^{171}Yb}$ is approximately two times less affected by the blackbody radiation shift in comparison with ${\rm ^{87}Sr}$~\cite{Porsev06}. Last but not least, ${\rm ^{171}Yb}$ has a much simpler Zeeman structure of the lasing states than ${\rm ^{87}Sr}$, which may considerably simplify the development of a repumping scheme.

The polarization of the optical lattice, as well as the magnetic field at the position of the atomic ensemble can be controlled to some extent only, and may contribute significantly to the uncertainties and fluctuations of the output frequency. The reasons are the polarization-dependent differential light shift and the Zeeman shift of the clock states. Note that most of these shifts, namely the vector light shift and the linear Zeeman shift, are proportional to the magnetic quantum number $m_F$ of the respective state. In passive optical clocks the error related to these effects can be suppressed by alternating between preparation of atoms with opposite Zeeman states $m_F=F$ and $m_F=-F$ in different interrogation cycles with subsequent averaging~\cite{Lemke09, Bloom14}. However, a direct application of this technique to the active optical lattice clocks seems to be impossible. At the same time, in active clocks it may be possible to pump the atoms into a balanced mixture of the upper lasing state sublevels with opposite $m_F$, and directly obtain lasing at the averaged frequency. A too large Zeeman splitting of the lasing transitions may destroy the synchronization between the two transitions, similarly to the loss of synchronization between different atomic ensembles~\cite{Xu14, Weiner15}. In this section we investigate such balanced lasing for spin-1/2 atoms. We introduce a differential Zeeman shift between lasing transitions with opposite $m_F$ into our model, and study the influence of this shift on the steady-state solutions and their stability. 

This section consists of 3 subsections. In the first one we specify our semiclassical model introduced in section \ref{sec:gmod} to the case of inhomogeneously broadened, incoherently pumped spin-1/2 atoms. In the second one we discuss possible stationary solutions. In the last subsection we perform the stability analysis, and discuss the main results.

\subsection{Specification of the model for spin-1/2 atoms.}
\label{sec:YbMod}

We consider an ensemble of active (inverted) atoms with total angular momentum $F=1/2$ in both the lower and upper lasing states, experiencing an external magnetic field causing a differential Zeeman shift of the atomic transitions. The atoms are coupled to a $\pi$-polarized cavity mode; for the sake of simplicity we suppose that the coupling coefficients $\g_{ge}$ are the same for all the atoms. Each $j$th atom has a detuning $\Delta_j$ from the cavity eigenfrequency $\omega_c$, caused by some external reason whose nature is not specified here. We suppose that individual atomic detunings $\Delta_j$ obey a normal distribution with zero detuning and dispersion $\Delta_0$:
\begin{equation}
\frac{dN}{d\Delta_j}=\frac{N}{\sqrt{2\pi}\Delta_0}\exp\left[-\frac{\Delta_j^2}{2\Delta_0^2}\right]. \label{e:33}
\end{equation}
Here $N$ is the total number of active atoms. Finally, all the atoms are incoherently repumped with the same rate $w$.

\begin{figure}
\begin{center}
\resizebox{0.45\textwidth}{!}
{\includegraphics{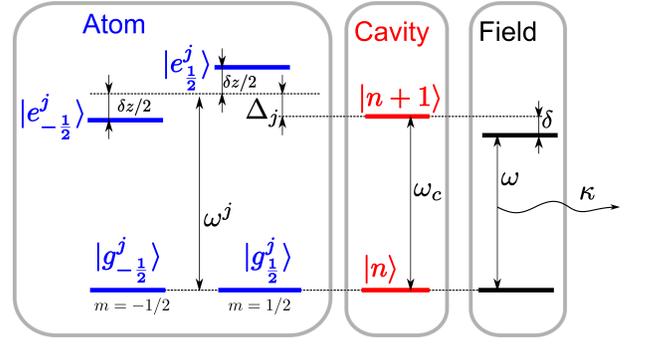}}
\end{center}
\caption{(color online) Structure of levels of individual atoms and notation for levels, frequencies and detunings: $\delta = \omega_c-\omega$, $\Delta_j=\omega^j-\omega_c$. Because only $\pi$-transitions can be excited by the cavity field, the relative energy shift between the levels $|g^j_{-1/2}\rangle$ and $|g^j_{1/2}\rangle$ is not significant.}
\label{fig:f2}
\end{figure}
Using the notation introduced in Figure~\ref{fig:f2}, we can write the Hamiltonian of the $j$th atom interacting with the cavity field in the corresponding rotating frame as
\begin{align}
\HH_a^{(j)} + \HH_{af}^{(j)}=\, & \hbar \sum_{m=-1/2}^{1/2}\left[\sig_{ee}^{j,m}(\delta+\Delta_j+\delta_z\, m)  \vphantom{\frac{\g_k}{2}} \right. \nonumber \\
 & + \left. \frac{\g_m}{2}\left(\a^+\sig^{j,m}_{ge}+\sig^{j,m}_{eg}\a \right) \right]. \label{eq:27}
\end{align}
Here $\sig^{j,m}_{\alpha,\beta}=|\alpha^j_m\rangle \langle \beta^j_m|$ ($\alpha,\beta \in \{e,g\}$), $\delta_z$ is a differential Zeeman splitting. the coupling coefficient $\g_m$ can be expressed as $\g_m=\g\,C^{Fm}_{Fm10}$, where $C^{Fm}_{Fm10}$ is a Clebsch-Gordan coefficient. To describe incoherent pumping~\cite{Meiser09} and spontaneous relaxations of the $j$th atom, we use the following Liouvillian superoperator 
\begin{align}
\LL_{j}[\hro]= & \sum_{l,n=-1/2}^{1/2}
 \left( \frac{\gamma_{ln}}{2}\left[2 \sig_{ge}^{j,nl}\hro  \sig^{j,ln}_{eg}-\sig^{j,ll}_{ee}\hro-\hro \sig^{j,ll}_{ee}  \right] \right. \nonumber \\
 &+ \left. \frac{w_{nl}}{2} \left[2 \sig_{eg}^{j,ln}\hro \sig^{j,nl}_{ge} -\sig^{j,nn}_{gg}\hro-\hro \sig^{j,nn}_{gg}  \right] \right),  \label{eq:28}
\end{align}
where $\sig^{j,nl}_{\alpha\beta}=|\alpha^j_n\rangle\langle \beta^j_l|$, $\gamma_{ln}=\gamma |C^{Fl}_{Fn1q}|^2$ (here $q=l-n$) is the rate of spontaneous decay from the state $|e^j_l\rangle$ to $|g^j_n\rangle$, $w_{nl}$ is the incoherent pumping rate from the state $|g^j_n\rangle$ to $|e^j_l\rangle$. For the sake of definiteness, we suppose that the atomic magnetic states are totally mixed during the repumping process:
\begin{equation}
w_{\pm\frac{1}{2},\pm\frac{1}{2}}=w_{\pm\frac{1}{2},\mp\frac{1}{2}}=\frac{w}{2}. \label{e:68}
\end{equation}
Also, we neglected here the incoherent dephasing rate (Rayleigh scattering in \cite{Bohnet14}).

The system is governed by the Born-Markov master equation (\ref{equ:1}), where all the components are defined in (\ref{equ:2}) -- (\ref{eq:7}) and (\ref{eq:27}) -- (\ref{e:68}). Now one can easily obtain the explicit form of the semiclassical equations (\ref{eq:8}) -- (\ref{eq:10}). 

\begin{figure*}
\begin{center}
\resizebox{1\textwidth}{!}
{\includegraphics{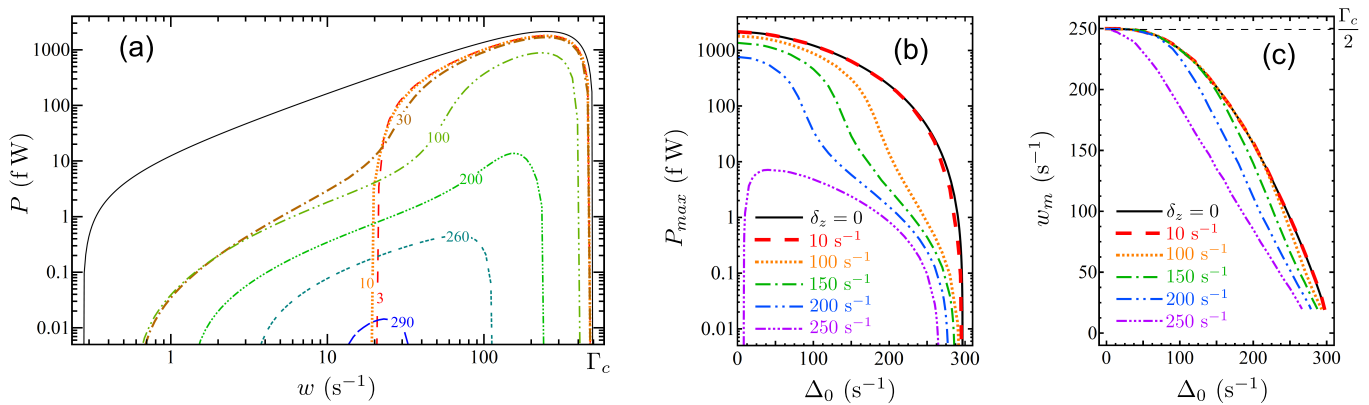}}
\end{center}
\caption{(color online) (a): Steady-state output power  versus the incoherent pumping rate $w$ for $\delta_z=100$ (style/coloured curves labelled by the values of $\Delta_0~{\rm (s^{-1})}$). The solid black curve, representing the output power of the laser with $\Delta_0=\delta_z=0$, is given as a reference. (b): Peak output power $P_{max}$ versus $\Delta_0$ for different values of $\delta_z$. (c): Repumping rate $w_m$ maximising the output power versus $\Delta_0$ for different values of $\delta_z$. Dashed horisontal line indicates $w_m=\Gamma_c/2$ at $\Delta_0=\delta_z=0$, see \cite{Meiser09} for details.}
\label{fi:f3}
\end{figure*}

For the numerical analysis, we partition the atoms into a number of groups (the graining of this partitioning has to be chosen fine enough, as described in section~\ref{sec:practLSA}). Then we identify the steady-state solutions of the semiclassical equations, and analyse their stability using the method presented in Section~\ref{sec:lsa}. 

For the sake of definiteness, we take the following parameters of the atomic ensemble and the cavity: number of atoms $N=10^5$, coupling coefficients $\g_{\pm 1/2}= \pm 50~{\rm s^{-1}}$, decay rate of the cavity field $\kappa=5 \times 10^5~{\rm s^{-1}}$. These parameters seem to be realistic (see also experiment~\cite{Norcia16} with ${\rm ^{87}Sr}$), and correspond to $\Gamma_c = N |\g_{ge}|^2/\kappa=500~{\rm s^{-1}}$. The parameter $\Gamma_c$ sets the upper lasing threshold (see Appendix \ref{sec:a1} for details), and its value will be kept constant throughout the paper. Also we have taken characteristic atomic parameters of ${\rm ^{171}Yb}$, namely the transition frequency $\omega=2\pi \times 518.3~{\rm THz}$, and the total spontaneous decay rate $\gamma=2\pi \times 43.5~{\rm mHz}$~\cite{Porsev04}.

\subsection{Possible steady-state solutions}

The system described in the previous subsection may have several steady-state solutions. The first one is a ``trivial'' zero-field solution. The second one is a non-trivial center-line solution which correspond at $\delta_z=0$ to the non-zero field solution of the two-level model, see Appendix~\ref{sec:appB} for details. The steady-state output power $P=\kappa \hbar \omega |\langle \a \rangle_{cw}|^2$ corresponding to this solution is shown in Figure~\ref{fi:f3}~(a) as a function of the repumping rate $w$ for $\delta_z=100~{\rm s^{-1}}$ and for various values of $\Delta_0$. Also we indicate the maximum output power $P_{max}$ (Figure~\ref{fi:f3}~(b)), and the pumping rate $w_m$ maximizing this power (Figure~\ref{fi:f3}~(c)) versus the inhomogeneous broadening $\Delta_0$ for different values of $\delta_z$. 

One can see that if the inhomogeneous broadening $\Delta_0$ and the differential Zeeman splitting $\delta_z$ are small in comparison with $\Gamma_c$ (5 or more times less), the optimized value of the output power and optimal repumping rate remain practically the same as for the system without any broadening and splitting. An increase of $\Delta_0$ and/or $\delta_z$ leads first to a decrease of the output power, and then to the disappearance of the center-line solution.

If $\delta_z\neq 0$, additional frequency-detuned ($\delta \neq 0$) steady-state solutions of the semiclassical equations may appear. These solutions correspond to the situation when lasing occurs primarily on one transition of the active atoms (with $m_F=1/2$ or $m_F=-1/2$), whereas the other transition pulls the field detuning $\delta$ towards the line-center position. These solutions appear in pairs, with the same field amplitudes and opposite detunings $\pm \delta$.


\begin{figure*}
\begin{center}
\resizebox{0.97\textwidth}{!}
{\includegraphics{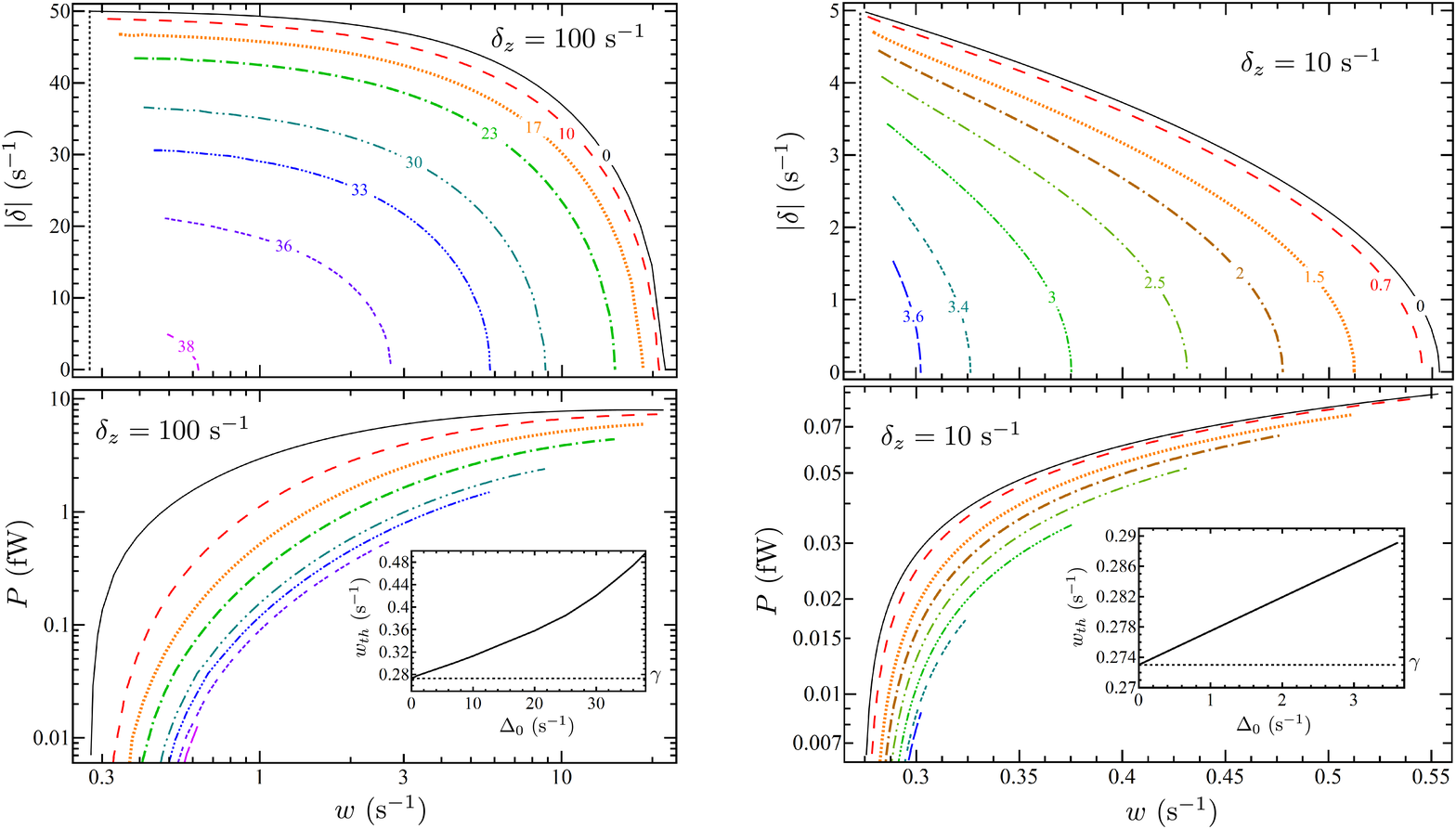}}
\end{center}
\caption{(color online) Frequency detunings $\delta=\omega-\omega_c$ (top) and output powers $P=\hbar \omega |\langle \a \rangle |^2 \kappa$ (bottom)  corresponding to the frequency-detuned solutions for two different values of $\delta_z$ (left and right pairs of plots) and different values of $\Delta_0$ versus the pumping rate $w$ (x-axis). Curves $\delta(w)$ are labelled by the values of $\Delta_0~{(s^{-1})}$, the same style-color encoding is valid for the power plot with the same $\delta_z$. Insets: lower threshold $w_{th}$ to lasing as a function of $\Delta_0$. Vertical dotted lines on the upper plots and horizontal dotted lines on the insets represent the fundamental lower threshold level $w=\gamma$.}
\label{fi:f4}
\end{figure*}

The output powers and frequency detunings corresponding to these frequency-detuned solutions are illustrated in Figure~\ref{fi:f4}. One can see that these solutions exist in a quite limited  range of $w$ only, and the lower (upper) limit of this range increases (decreases) with increasing (decreasing) $\Delta_0$ respectively. Also, the upper limit grows with an increase of the Zeeman splitting $\delta_z$.  On the lower limit $w_{th}$ (coinciding with the lower threshold to lasing), the detuning $\delta$ is maximal; it decreases with increasing $w$ until it reaches zero. Essentially, on the upper limit, both detuned solutions merge. 

\subsection{Analysis of stability}

\begin{figure*}[ht]
\begin{center}
\resizebox{1\textwidth}{!}
{\includegraphics{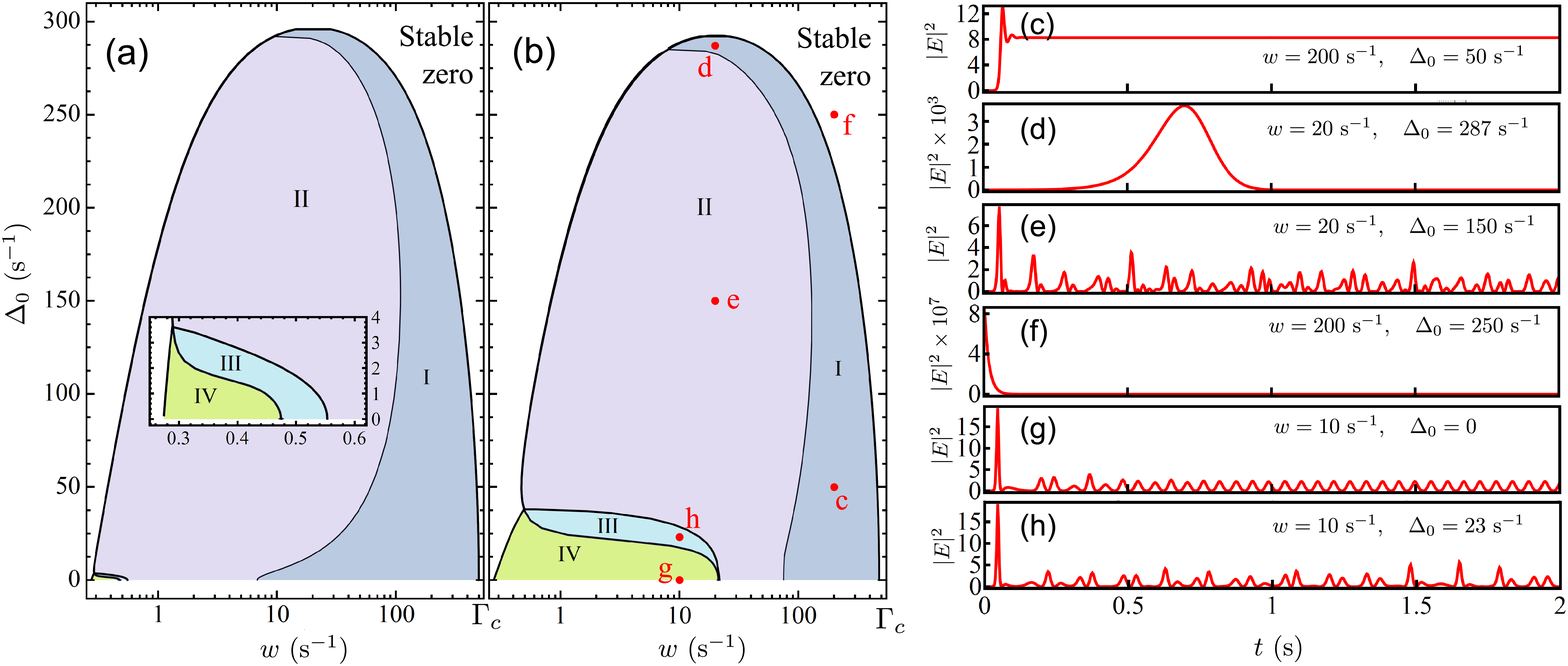}}
\end{center}
\caption{(color online) (a) -- (b): domains of existence and stability of various steady-state solutions in the $(w,\Delta_0)$ plane for $\delta_z=10~{\rm s^{-1}}$ (a), and $\delta_z=100~{\rm s^{-1}}$ (b).  A zero-field solution exist everywhere, but is stable only in the unshaded domain, where no other solution exists. Other domains are: I: line-centered solution exists and is stable; II: line-centered solution exists but is instable; III: line-centered and detuned solutions exist but are instable, IV: detuned solutions exist but are instable. The inset is an enlarged view of the area near the lower lasing threshold for $\delta_z=10~{\rm s^{-1}}$. (c) -- (h): time evolution of the intracavity photon number $|E|^2$  according the numerical simulation for $\delta_z=100~{\rm s^{-1}}$ and different values of $w$ and $\Delta_0$ (given as plot labels; also indicated as red dots in (b) with respective labels c -- h). The seed value of $|E|^2$ was set to $10^{-6}$ for this simulation.}
\label{fi:f5}
\end{figure*}

We evaluate the stability of the non-zero field steady-state solutions (both center-line and detuned) using a full numerical procedure, i.e. by partitioning the atoms into a number of groups (identical to the one used for the search of steady-state solutions), building the function $\DD(\lambda)$ according to (\ref{e:30}), and tracing its argument along the imaginary axis, as described in Section~\ref{sec:practLSA}. 

For the zero-field solution, we build the function $\DD(\lambda)$ explicitly. First let us give the explicit expressions for the components of equations (\ref{eq:8}) -- (\ref{eq:10}). We use the following notation for the single-atom state vector (normalization condition is taken into account): 
\begin{widetext}
\begin{equation}
\overline{\langle \sig^i \rangle}= \left( 
\begin{array}{ccccccc}
\langle \sig^{j,-1/2}_{ee} \rangle\, , &
\langle \sig^{j,-1/2}_{ge} \rangle\, , &
\langle \sig^{j,-1/2}_{eg} \rangle\, , &
\langle \sig^{j,1/2}_{gg} \rangle\, , &
\langle \sig^{j,1/2}_{ee} \rangle\, , &
\langle \sig^{j,1/2}_{ge} \rangle\, , &
\langle \sig^{j,1/2}_{eg} \rangle 
\end{array}
\right)^T, \label{e:69}
\end{equation}
where the superscript $T$ denotes transposition. 
Then the matrices $\AAA^{(j)}$, $\GGG^{(j)}$, and the constant term $\BBB$ are:

\begin{equation}
\AAA^{(j)}=\left(
\begin{array}{ccccccc}
-\frac{w}{2}-\gamma & -\frac{i E^* \g}{2 \sqrt{3}} & \frac{i E \g}{2 \sqrt{3}} & 0 & -\frac{w}{2} & 0 & 0 \\\\
-\frac{i E \g}{\sqrt{3}} & -\frac{w+\gamma}{2}-i\Delta_j^- & 0 & -\frac{i E \g}{2 \sqrt{3}} & -\frac{i E \g}{2 \sqrt{3}}  & 0 & 0 \\\\
\frac{i E^* \g}{\sqrt{3}} & 0 & -\frac{w+\gamma}{2}+i\Delta_j^- & \frac{i E^* \g}{2 \sqrt{3}} & \frac{i E^* \g}{2 \sqrt{3}} & 0 & 0 \\\\
\frac{2 \gamma}{3} & 0 & 0 & -w & \frac{\gamma}{3} & -\frac{i E^* \g}{2 \sqrt{3}} & \frac{i E \g}{2 \sqrt{3}} \\\\
-\frac{w}{2} & 0 & 0 & 0 & -\frac{w}{2}-\gamma & \frac{i E^* \g}{2 \sqrt{3}} & -\frac{i E \g}{2 \sqrt{3}} \\\\
0 & 0 & 0 & -\frac{i E \g}{2\sqrt{3}} & \frac{i E \g}{2\sqrt{3}} & -\frac{w+\gamma}{2}-i\Delta_j^+ & 0 \\\\
0 & 0 & 0 & \frac{i E^* \g}{2\sqrt{3}} & -\frac{i E^* \g}{2\sqrt{3}} & 0 &-\frac{w+\gamma}{2}+i\Delta_j^+
\end{array}
\right),  \label{e:70}
\end{equation}
\end{widetext}
\begin{equation}
\GGG^{(j)}=\left( 
\begin{array}{ccccccc}
0 & \frac{i \g}{2\sqrt{3}} & 0 & 0 & 0 & -\frac{i \g}{2\sqrt{3}} & 0 \\
0 & 0 & -\frac{i \g}{2\sqrt{3}} & 0 & 0 & 0 & \frac{i \g}{2\sqrt{3}} 
\end{array}
\right),  \label{e:72}
\end{equation}
\begin{equation}
\BBB^{(j)}=\left( \begin{array}{ccccccc}
\frac{w}{2}\, ,&
\frac{i E \g}{2\sqrt{3}} \, ,&
-\frac{i E^* \g}{2\sqrt{3}} \, ,&
0 \, ,&
\frac{w}{2} \, ,&
0 \, ,&
0 
\end{array}\right)^T,  \label{e:72}
\end{equation}
where $\Delta_j^\pm=\Delta_j\pm\delta_z/2$, $E=\langle \a \rangle$, $E^*=\langle \a^* \rangle$. 

To find the zero-field solution, one needs to set $E=E^*=0$ in (\ref{e:70}), (\ref{e:72}), and solve (\ref{eq:10}). We obtain
\begin{equation}
\SSS^{j}=\left( 
\begin{array}{ccccccc}
\frac{w}{2(w+\gamma)} ,&0,  &0, & \frac{\gamma}{2(w+\gamma)}, &  \frac{w}{2(w+\gamma)} , & 0 , & 0
\end{array}
\right)^T.  \label{e:73}
\end{equation}

Using decompositions (\ref{eq:13}), (\ref{eq:14}), definitions (\ref{eq:18}), (\ref{eq:20}), and taking the integral in (\ref{e:47}) using distribution (\ref{e:33}), we obtain for the zero-field solution
\begin{equation}
\DD(\lambda)=\left(\frac{\lambda+\kappa/2-\MM}{\lambda+\kappa/2}\right)^2, \label{e:74}
\end{equation}
where
\begin{align}
\MM(\lambda)&=\frac{\g^2 (w-\gamma)}{12 (w+\gamma)}\times \\
& \left[
\frac{\eta_1}{8\Delta_0^2 \sqrt{\pi}}\, \Theta\left(\frac{\eta_1^2}{8 \Delta_0^2} \right)
+\frac{\eta_2}{8\Delta_0^2 \sqrt{\pi}}\, \Theta\left(\frac{\eta_2^2}{8 \Delta_0^2} \right)
\right]. \label{e:75}
\end{align}
Here the function $\Theta$ is defined via the complementary error function as
\begin{equation}
\Theta (y)= \int_{-\infty}^{\infty}\frac{e^{-x^2}dx}{x^2+y}= \frac{e^y \, \pi\,  {\rm erfc} (\sqrt{y})}{\sqrt{y}}, \label{e:43}
\end{equation}
and 
\begin{equation}
\eta_{1,2}=w+\gamma+2 \lambda \pm i \delta_z. \label{e:76}
\end{equation}

In Figure~\ref{fi:f5} we present domains of existence and stability of different steady-state solutions for $\delta_z=10~{\rm s^{-1}}$ (a) and $\delta_z=100~{\rm s^{-1}}$ (b). One can see that these diagrams resembles the ones obtained for two-level atoms (see Figure~\ref{fig:B2}~(a) in Appendix~\ref{sec:appB}) everywhere, except an area near the origin, where $w$ and $\Delta_0$ are smaller than $\delta_z$. Also we should note that frequency-detuned solutions are always instable. The domain of stability of the zero-field solution coincides with the complement of the domain of existence of any non-zero steady-state solution, i.e. the zero-field solution is stable, if and only if no non-zero field solutions exists. 

We also present the time evolution of the mean intracavity photon number $|E|^2$ for selected values of $w$ and $\Delta_0$ at $\delta_z=100~{\rm s^{-1}}$ in Figure~\ref{fi:f5} (c) -- (h). Note that in the instable regimes, the photon number may demonstrate either irregular chaotic behavior, like in Figure~\ref{fi:f5} (e), or regular pulsation, like in Figure~\ref{fi:f5} (g). This pulsation might be interpreted as independent lasing on two transitions, if the frequency of this pulsation would be equal to $\delta_z=100~{\rm s^{-1}}\approx 2\pi \times 15.9~{\rm Hz}$. However, this pulsation frequency is higher (about 18 Hz), and slightly grows with increasing $w$.

Another remarkable fact is that an increase of the pumping parameter $w$ is accompanied by a transition from an instable to a stable lasing regime. This behaviour differs from the one described in~\cite{Haken, Abraham85, Zhang85, Mandel85}, where it has been shown that instability appears only if the pumping rate exceeds some ``second laser threshold''. We found that the reason for this inversion is that in our model the total decoherence rate $\gamma_{\perp}=(w+\gamma)/2$ is primarily determined by the repumping rate $w$. Therefore, increasing $w$ leads to an increase of the {\em homogeneous} broadening and a suppression of the fluctuations of the cavity field. In contrast, in~\cite{Haken, Abraham85, Zhang85, Mandel85} the authors introduced pumping and relaxation rates as independent parameters. We should note that introducing an additional inhomogeneous dephasing leads to the stabilization of the lasing near the lower lasing threshold, in correspondence with~\cite{Haken, Abraham85, Zhang85, Mandel85}, see  Appendix~\ref{app:B2} for details.

In general, we can conclude that a stable lasing regime with high output power can be attained, if both the inhomogeneous broadening parameter $\Delta_0$ and the differential Zeeman shift $\delta_z$ are at least a few times smaller than the incoherent repumping rate $w$, which is limited by the upper lasing threshold $\Gamma_c$.
\section{Outlook}
\label{sec:outlook}
Here we briefly review the obtained results and discuss some perspectives of building an active optical frequency standard using inhomogeneously broadened ensembles and simultaneous lasing on different transitions interacting with the same cavity mode.
\subsection{Optical lattice clocks with compensated first-order Zeeman and vector light shifts}

In the previous section we investigated the optical lattice laser with an inhomogeneously broadened ensemble of incoherently pumped alkali-earth-like atoms with total angular momentum $F=1/2$ in both the upper and lower lasing states, such as $^{171}$Yb, $^{199}$Hg, ${\rm ^{111}Cd}$ and ${\rm ^{113}Cd}$. We considered the situation when both $\pi$-polarized lasing transitions are pumped equally, and an differential Zeeman shift $\delta z$ is present. We found that, as long as both the inhomogeneous broadening parameter $\Delta_0$ and the differential Zeeman shift $\delta_z$ are small in comparison with the pumping rate $w$, their influence on the output power and stability of the lasing regime remains minor. In other words, if $\Delta_0, \delta_z \ll w$, one can neglect inhomogeneous broadening and Zeeman splitting for the description of the bad cavity laser, and if $\Delta_0, \delta_z \ll \Gamma_c = N g^2/\kappa$, the optimum regime and maximum output power will be similar to the one for two-level lasers without inhomogeneous broadening. 

This finding opens the possibility to build an active optical frequency standards using inhomogeneously broadened ensembles of atoms, and to suppress the linear Zeeman and vector light shifts by means of balanced lasing on the transitions between the pairs of the upper and lower lasing states with opposite $m_F$. We should recall, however, that if the imhomogeneous width and/or differential Zeeman shift $\delta z$ occur to be of order of or larger than the decoherence rate, the stability may be lost, and/or the output laser power may be significantly reduced, because most of the atoms will be far from resonance with the cavity field.

Also we checked the robustness of the center-line solution with respect to an imbalance in the repumping rates caused, for example, by a slight ellipticity of the repumping fields. We implemented an imbalanced repumping rate in the form $w_{m,\pm 1/2}=\frac{w}{2} (1 \pm \epsilon)$ (where $m$ is the magnetic quantum number of the ground state), and we obtained that the frequency shift of the output radiation $\delta \approx \delta_z \epsilon$, if  $w/\Gamma_c$ lies between 0.2 and 0.9. Therefore, the uncertainty introduced by the first-order Zeeman and vector light shifts remains, but can be suppressed by the remaining pumping imbalance $\epsilon$, in comparison with the lasing on only one of the possible lasing transitions.

\subsection{Active optical clocks based on large ion crystal}

The results outlined above open up another possibility for implementing an active optical frequency standards. Namely, such a standard can be realized with Coulomb crystals formed by ions trapped in RF Paul (or Penning) traps. The main advantage of such an approach is the long lifetime of ions in the trap, which absolves the experimentalist from the need for sophisticated methods to compensate for atom losses. 

Up to now, ion optical clocks have been built primarily using single ions or small few-ion ensembles~\cite{Keller16}; large ensembles have not been used because of micromotion-related second-order Doppler, Stark, and quadrupole shifts causing significant inhomogeneous broadening. However, these limitations may, in principle, be overcome for some ion species~\cite{Arnold15}.

It appears to be possible to build a bad-cavity laser on ions trapped in a linear Paul trap, if the lasing transition fulfills some specific requirements. First, this transition should be strong enough to realize the strong coupling regime, and should lie in a convenient wavelength region, where it is possible to build a high-finesse cavity. Second, efficient cooling and pumping into the upper clock state should be possible. Third, the lasing states should have negative differential polarizability $\Delta_\alpha=\alpha_e-\alpha_g$. This allows to compensate (in leading order) the micromotion-induced second-order Doppler shift and the Stark shift at a so-called {\em magic} frequency 
\begin{equation}
\Omega_{0}=\frac{q}{mc}\sqrt{\frac{\hbar \omega}{-\Delta \alpha}}
\end{equation}
of the RF trapping field. Here $q$ and $m$ are the charge and the  mass of the ion, $\omega$ is the frequency of the clock transition.
\begin{figure}
\begin{center}
\resizebox{0.4\textwidth}{!}
{\includegraphics{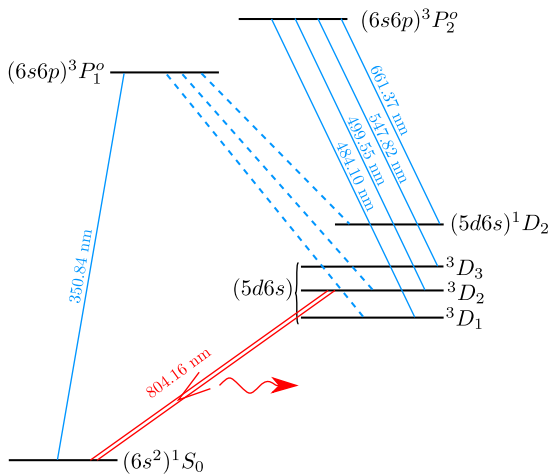}}
\end{center}
\caption{(color online) General pumping scheme (hyperfine structure not shown) for a 804 nm bad cavity laser on {$^{176}$Lu} ions. Dashed lines denote the most relevant spontaneous decays, solid lines correspond to both spontaneous and laser-induced transitions (wavelengths are indicated).}
\label{fig:Lu}
\end{figure}

The combination of these properties can be found, for example, in the ${^3D_2} \rightarrow {^1S_0}$ transition in ${\rm ^{176}Lu^+}$ ions. A detailed analysis of such a system will be published soon~\cite{We}, here we only briefly mention the main concepts. A possible repumping scheme is shown in Figure~\ref{fig:Lu}: a 350.84~nm pumping laser populates the $^3P_1^o$ state which decays with a 42~\% probability into the $^3D_2$ upper lasing state~\cite{Paez16}. To pump the ions from the  ${^3D_1}$ and ${^1D_2}$ states into ${^3D_2}$, three additional lasers are required: 661.37~nm, 547.82~nm, and 484.10~nm. Because the nucleus of $^{176}$Lu has non-zero angular momentum $I=7$, these lasers should have several frequency components to cover the hyperfine structure of the $D$ states. Finally, a 5-component 499.55\,nm laser should be employed to pump the populations into the upper lasing state with specific $F=F_e$ and $m_F=0$. This can be realized, if one component of this laser is tuned in resonance with the $|^3D_2, F_e\rangle \rightarrow |^3P_2^o, F_e \rangle$ transition and polarized along the $z$ axis of the trap coinciding with the direction of the auxiliary magnetic field.

We consider a cold Coulomb crystal of $\rm Lu^+$ ions in a linear Paul trap, where the RF field lies in the $(x,y)$ plane orthogonal to the auxiliary magnetic field. Then the non-compensated oscillating electric field acting on the ions lies primarily in this plane. Also we suppose that the Zeeman splitting is large in comparison with the Stark shift. 

According to~\cite{Paez16}, the spontaneous rate of the lasing transition $\gamma= 4.19 \times 10^{-2}~{\rm s^{-1}}$, the differential scalar polarizability $\Delta \alpha_0 = -0.9~a_0^3$, and the tensor polarizability of the upper state $\alpha_2=-5.6~a_0^3$, where $a_0$ is the Bohr radius. Taking $|^3D_2,F_e=8,m_F=0 \rangle$ as the upper lasing state, we can find the magic frequency $\Omega_0=2\pi \times 45.5~{\rm MHz}$ following the method described in~\cite{Itano00}. 

For an estimation of $\Gamma_c$ we suppose that the trap is spherically-symmetric with a pseudopotential oscillation frequency $\omega_z=2\pi \times 2~{\rm MHz}$, and contains $10^5$ ions. With the cavity waist being equal to the radius of this Coulomb crystal (about $\rm 80\,\mu m$), and the cavity finesse $\mathcal{F}=10^5$, we find $\Gamma_c \approx 130~{\rm s^{-1}}$, whereas the remaining broadening due to higher-order contributions from the Stark and second-order Doppler shifts will be about $20~{\rm s^{-1}}$. Therefore, the condition $\Delta \ll \Gamma_c$ will be fulfilled, and a trapped-ion bad-cavity laser on this transition operating in a stable regime seems to be realistic. We can increase $\Gamma_c$ further using a cigar-shaped trap instead of a spherical one. 

Of course, there is a strong gap between the idea of a bad cavity laser and the scheme of an active optical clock, where different factors deteriorating the performance should be considered and minimized. A detailed study of these effects lies beyond the scope of the present paper.


\section{Conclusion}
In this paper we introduced a new method for a numerical linear stability analysis of inhomogeneously broadened running-wave lasers or lasers where the active atoms are confined in space (like the optical lattice laser). Our method consists in tracing the argument of a specific function over the imaginary axis in the complex plane. Both computational and memory costs of this method are {\em linear} in the number of partitions, which allows us to perform extended studies of the stability of lasers with complex multilevel gain atoms and inhomogeneous broadening within a wide range of parameters. 

Using this method, we investigated the stability of the optical lattice laser with an inhomogeneously broadened ensemble of incoherently pumped alkali-earth-like atoms with total angular momentum $F=1/2$ in both the upper and lower lasing states, such as $^{171}$Yb, $^{199}$Hg, ${\rm ^{111}Cd}$ and ${\rm ^{113}Cd}$. The situation in which both $\pi$-polarized lasing transitions are pumped equally, while a differential Zeeman shift is present, has been considered. We investigated possible steady-state solutions, and conditions for their existence and stability. We found that stable lasing and high output power can be attained, if both the inhomogeneous broadening parameter $\Delta_0$ and the differential Zeeman shift $\delta_z$ are small in comparison with the pumping rate $w$. Increasing the inhomogeneous broadening and/or differential Zeeman shift will partially suppress the lasing, and may eventually destroy the stability. Also, we showed that if $\Delta_0$ and $\delta_z$ are both small (5 or more times less) in comparison with $\Gamma_c = N g_{eg}^2/\kappa$, then the maximum output power $P_{max}$ and the optimal pumping rate $w_m$ maximizing this output power will be close to the ones predicted by a simple two-level model~\cite{Meiser09}, and the laser will operate in a stable regime with these values. 

This fact allows to use balanced lasing on two $\pi$-polarized lasing transitions between pairs of states with opposite values of $m_F$ for the suppression of the first-order Zeeman and the vector light shift in optical lattice lasers. Also, it seems to be possible to build a bad cavity laser (and probably an active optical clock) on multi-ion ensembles trapped in axial Paul traps. This technique may be helpful to avoid sophisticated methods to compensation losses because of the long lifetime of the ions in the trap.

\section{Acknowledgements}
This study has been supported by the FWF project I~1602 and the EU-FET-Open project 664732 NuClock.

\appendix
\section{Stability of the steady-state solution for the two-level model with incoherent pumping}
\label{sec:a1}
Here we briefly overview the instabilities arising in a two-level bad cavity laser without inhomogeneous broadening. Although this system has been considered in textbooks~\cite{Haken}, it is useful to review it using the notations introduced in~\cite{Meiser09} and subsequent publications~\cite{Bohnet14,Xu13,Xu14,Xu15}. 

We start from the Born-Markov master equation for the reduced atom-field density matrix $\hro$. For the sake of simplicity, we assume the cavity mode to be exactly in resonance with the atomic transition. Then the density matrix is governed by the equation (\ref{equ:1}), where the Hamiltonian $\HH$  after the transformation (\ref{eq:5}) becomes
\begin{equation}
\HH=\frac{\hbar \g}{2}\sum_{j=1}^N(\sig^j_+\a +\a^+\sig^j_-). \label{eq:A1}
\end{equation}
Here $\sig^{j}_{+}=\sig^{j}_{eg}$, $\sig^{j}_{-}=\sig^{j}_{ge}$.
The single-atom Liouvillian $\LL_j$ is
\begin{align}
\LL_j=&\frac{\gamma}{2} \left(2\sig^j_-\hro\sig^j_+-\sig^j_{ee}\hro-\hro\sig^j_{ee}\ \right)+\frac{\gamma_R}{2} \left(\sig^j_z\hro \sig^j_z-\hro \right)\nonumber \\
&+\frac{w}{2}\left(2\sig^j_+ \hro\sig^j_- -\sig^j_{gg}\hro-\hro\sig^j_{gg}\ \right), \label{A:2}
\end{align}
where $\gamma$ is the rate of spontaneous decay of the lasing transition, $w$ is the rate of incoherent pumping, $\gamma_R$ is the incoherent dephasing rate, $\sig^j_z=\sig^j_{ee}-\sig^j_{gg}$. The Liouvillian of the cavity field is given by (\ref{equ:2}). Introducing macroscopic variables
\begin{align}
E=\langle \a \rangle, \quad p= -i\sum_{j=1}^N \sig^j_-, \quad D= \sum_{j=1}^N \sig^j_z,
\label{eq:A3}
\end{align}
we can write the semiclassical equations as
\begin{align}
\dot{E}&=-\frac{\kappa}{2} E+\frac{\g}{2}p \label{eq:A4}\\
\dot{p}&=-\gamma_{\perp} p+\frac{\g}{2}DE \label{eq:A5}\\
\dot{D}&=\gamma_{\parallel} (D_0-D)-\g (E p^*+E^*p). \label{eq:A6}
\end{align}
Here $\gamma_{\parallel}=(w+\gamma)$, $\gamma_{\perp}=(w+\gamma)/2\gamma_R$, $D_0=N d_0=N(w-\gamma)/(w+\gamma)$. The non-zero steady-state solution (indexed by ``cw'') is
\begin{align}
p_{cw}&= e^{i\phi} \sqrt{\frac{\kappa \gamma_{\parallel}}{2 \g^2} \left( 
D_0-\frac{2 \gamma_{\perp} \kappa}{\g^2},
\right)} \nonumber \\
D_{cw}&=2\gamma_{\perp}\kappa/\g^2, \quad E_{cw}=p_{cw}\,  \g/ \kappa,  \label{eq:A7}
\end{align}
where $\phi$ is an arbitrary phase. These solutions exist only if 
\begin{align}
w < \frac{N d_0 \g^2}{\kappa}-\gamma-{2\gamma_R}. \label{eq:A8}
\end{align}
If $\gamma \ll \Gamma_c$, this condition can be rewritten as:
\begin{align}
\gamma \frac{1+2 \gamma_R/\Gamma_c}{1-2 \gamma_R/\Gamma_c} < w < \Gamma_c - 2 \gamma_R. \label{e:A9}
\end{align}
We refer to these limits as the {\em lower} and the {\em upper} laser thresholds, following~\cite{Meiser09}. 

While performing the linear stability analysis of the solution (\ref{eq:A7}), one can fix the phase $\phi=0$, following Haken~\cite{Haken}. It leads to the loss of two roots of the characteristic polynomial, but does not impair the stability analysis (one of the lost roots corresponding to the phase invariance being equal to zero, and another one corresponding to the decay of a phase imbalance between the atoms and the cavity mode always being negative). Introducing dimensionless variations
\begin{equation}
\varepsilon=\frac{E-E_{cw}}{E_{cw}}, \quad \varrho=\frac{p-p_{cw}}{p_{cw}}, \quad 
\vartheta=\frac{D-D_{cw}}{D_{cw}},  \label{eq:A9}
\end{equation}
one obtains the set of linearized equations
\begin{align}
\dot{\varepsilon}=&\, \frac{\kappa}{2} (-\varepsilon + \varrho), \nonumber \\
\dot{\varrho}=&\, \gamma_{\perp} (\varepsilon-\varrho+\vartheta),  \label{eq:A10} \\ 
\dot{\vartheta}=&\, - \gamma_{\parallel} [\Lambda(\varepsilon+\varrho)+\vartheta], \nonumber
\end{align}
where $\Lambda=(D_0/D_{cw}-1)$. To determine the stability domain, one can apply the Routh-Hurwitz criterion to the characteristic polynomial of (\ref{eq:A10}). The steady-state solution is stable, if
\begin{equation}
\gamma_{\perp} \Lambda \left(\frac{\kappa}{2}-\gamma_{\perp}-\gamma_{\parallel} \right)< 
\left(\frac{\kappa}{2}+\gamma_{\perp}+\gamma_{\parallel}\right)
\left(\frac{\kappa}{2}+\gamma_{\perp}\right) . \label{eq:A11}
\end{equation}
In other words, instability arises only when both conditions
\begin{align}
\kappa & > 2( \gamma_{\parallel}+\gamma_{\perp}),\quad {\rm and} \label{eq:A12} \\
\Lambda & > \frac{\left(\frac{\kappa}{2}+\gamma_{\perp}+\gamma_{\parallel}\right)
\left(\frac{\kappa}{2}+\gamma_{\perp}\right)}{\gamma_{\perp} \left(\frac{\kappa}{2}-\gamma_{\perp}-\gamma_{\parallel} \right)}. \label{eq:A13}
\end{align}
are fulfilled. From (\ref{eq:A13}) follows
\begin{equation}
N \g^2 - 2 \gamma_{\perp} \kappa > \kappa^2. \label{eq:A14}
\end{equation}
Taking $\kappa = 10^6~{\rm s^{-1}}>> \gamma_{\perp}$, $\g = 10^2~{\rm s^{-1}}$ (realistic parameters of a high-performance bad cavity laser on a ${^1S_0} \leftrightarrow {^3P_0}$ transition in alkali-earth-like atoms estimated in \cite{Kazakov14}), one finds that instabilities arise only when the total number of active atoms $N>10^8$. This value seems to be unrealistic in optical lattice laser systems. On the other hand, for a laser operating on the ${^1S_0} \leftrightarrow {^3P_1}$ transition, similar to the one presented in~\cite{Norcia15}, this condition is easily attainable because of the much stronger atom-cavity coupling. Note that in~\cite{Norcia15}, oscillations of the output power have been observed in a cavity-enhanced pulse, without optical pumping.

\section{Optical lattice laser with inhomogeneously broadened ensemble of incoherently pumped two-level atoms}
\label{sec:appB}

\begin{figure*}
\begin{center}
\resizebox{0.99\textwidth}{!}
{\includegraphics{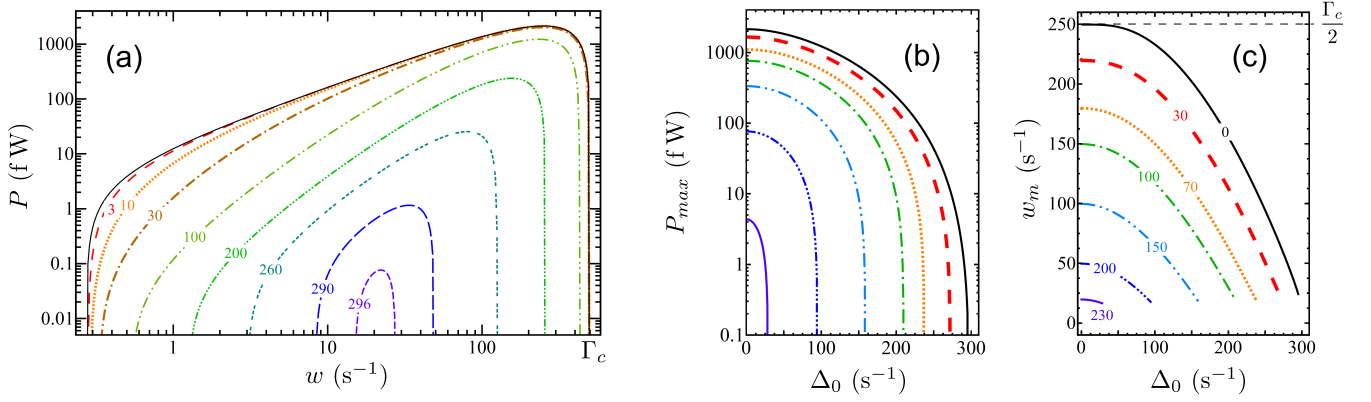}}
\end{center}
\caption{(color online) (a): Steady-state output power versus the incoherent repumping rate $w$ for atomic ensembles without (solid black curve) and with inhomogeneous broadening (colored curves labelled by values of $\Delta_0~{\rm (s^{-1})}$) at $\gamma_R=0$. (b) and (c): Peak output power $P_{max}$ (b) and repumping rate $w_m$ maximizing the output power (c) versus $\Delta_0$ for different values of $\gamma_R$. Curves $w_m(\Delta_0)$ are labelled by the values of $\gamma_R~({\rm s^{-1}})$; the same style-color encoding is valid for the plot of $P_{max}(\Delta_0)$ (b) with the same $\gamma_R$. Dashed horisontal line indicates $w_m=\Gamma_c/2$.}
\label{fig:B1}
\end{figure*}

Here we construct the function $\DD(\lambda)$ and perform the linear stability analysis for a laser with inhomogeneously broadened and incoherently pumped two-level active atoms. The aim of this Appendix is to illustrate the applicability of our method for analytical treatments, and to overview the most important characteristics of such a system. We limit our consideration to line-centred normally broadened distributions. Also, we will use here the notation introduced in~\cite{Meiser09} and subsequent theoretical papers~\cite{Bohnet14,Xu13,Xu14,Xu15}, to establish a link with modern studies of active optical frequency standards. 

We should also note that a 2-level system may be realized on ${^3P_0} \rightarrow {^1S_0}$ transitions in bosonic isotopes of alkaline-earth atoms. Such transitions may be slightly allowed in external magnetic fields~\cite{Taichenachev06}, or in circularly-polarized optical lattices~\cite{Ovsiannikov07}. This laser would require a simpler repumping scheme because of the absence of hyperfine splittings of intermediate levels used for the pumping. On the other hand, additional challenges may arise from the reduced strength of the lasing transition (at reasonable values of the magnetic or trapping fields), and collisions between the identical bosons.

\subsection{Semiclassical equations and steady-state solutions for the two-level model}
We start from the Born-Markov master equation for the reduced atom-field density matrix $\hro$. Then the density matrix is governed by the equation (\ref{equ:1}), where the Hamiltonian $\HH$  after the transformation (\ref{eq:5}) becomes
\begin{equation}
\HH=\hbar \sum_{j=1}^N\left[ \frac{\g}{2}(\sig^j_+\a +\a^+\sig^j_-) + \Delta_j \sig_{ee}^j \right]. \label{eq:B1}
\end{equation}
Here $\sig^{j}_{+}=\sig^{j}_{eg}$, $\sig^{j}_{-}=\sig^{j}_{ge}$, $\Delta_j=\omega_j-\omega_c$ is a detuning of the frequency $\omega_j$ of the lasing transition of the $j$th atom from the cavity mode frequency $\omega_c$. The single-atom Liouvillian $\LL_j$ is given by (\ref{A:2}), and the Liouvillian of the cavity field is given by (\ref{equ:2}).

We suppose that the detunings $\Delta_j$ of the atoms obey a normal distribution (\ref{e:33}) with zero mean (line-center operation) and dispersion $\Delta_0$. In such a case, the frequency $\omega$ of the laser radiation coincides with the mode eigenfrequency $\omega_c$. Choosing $\delta=0$ and introducing
\begin{equation}
\overline{\left\langle \sig^j \right \rangle}= \left(
\begin{array}{c}
\langle \sig^j_- \rangle \\ \langle \sig^j_+ \rangle \\ \langle \sig^j_z \rangle
\end{array} \right) \label{eq:B3}
\end{equation}
we can build the set of semiclassical equations of the form (\ref{eq:8}) -- (\ref{eq:10}), where $\delta=0$,
\begin{equation}
\AAA^{(j)}=\left(
\begin{array}{ccc}
-\gamma_\perp-i\Delta_j & 0 & \frac{i \g \langle \a \rangle}{2} \\
0 & -\gamma_\perp+i\Delta_j & - \frac{i \g \langle \a^+ \rangle}{2} \\
i \g \langle \a\rangle & - i \g \langle \a^+ \rangle & - \gamma_{\parallel}
\end{array}
\right), \label{eq:B4} \\
\end{equation}
\begin{align}
\BBB^{(j)} & =\left(
\begin{array}{c}
0 \\ 0 \\ w-\gamma
\end{array}
\right), \label{eq:B5} \\
\GG^{\prime (j)}&=\frac{i \g}{2}\left(
\begin{array}{ccc}
-1 & 0 & 0
\end{array}
\right), \label{eq:B6} \\
\GG^{\prime \prime (j)}&=\frac{i \g}{2}\left(
\begin{array}{ccc}
0 & 1 & 0
\end{array}
\right). \label{eq:B7}
\end{align}
Here $\gamma_{\perp}=(w+\gamma)/2+\gamma_R$ and $\gamma_{\parallel}=(w+\gamma)$. 

There are two possible steady-state solutions of equations (\ref{eq:8}) -- (\ref{eq:10}). The first one is a trivial zero-field solution:
\begin{align}
\langle \a \rangle = \langle \a^+ \rangle = 0; \quad \langle \sig^j_-\rangle = \langle \sig^j_+ \rangle = 0; \quad \langle \sig_z^j \rangle_0= d_0, \label{eq:B8}
\end{align}
where $d_0=(w-\gamma)/(w+\gamma)$. The second, non-trivial solution may be obtained after some algebra in the following form (see also \cite{Abraham85}):
\begin{align}
\langle \sig^j_- \rangle_{cw}& = \langle \sig^j_+ \rangle_{cw}^*=\frac{i \g \langle \a \rangle_{cw} \langle \sig^j_z \rangle_{cw}}{2 (\gamma_{\perp}+i \Delta_j)}, \label{eq:B9} \\
\langle \sig_z^j \rangle_{cw}&=\frac{(w-\gamma)(\gamma_{\perp}^2+\Delta_j^2)}{(w+\gamma)(\gamma_{\perp}^2+\Delta_j^2)+|\langle \a \rangle_{cw}|^2 \g^2 \gamma_{\perp}} , \label{eq:B10} \\
|\langle \a \rangle_{cw}|^2 & = \frac{2 \Delta_0^2 \gamma_{\parallel}}{\g^2 \gamma_{\perp}} \left[
\Theta^{-1}\left( \frac{4 \Delta_0^2 \kappa \sqrt{\pi}}{\g^2 N \gamma_{\perp} d_0}  \right) - \frac{\gamma_\perp^2}{2 \Delta_0^2}
\right], \label{eq:B11}
\end{align}
where the function $\Theta^{-1}$ is inverse to the function (\ref{e:43}).

The influence of the inhomogeneous broadening on the steady-state output power $P$ estimated as $P=\hbar \omega \kappa |\langle \a \rangle|^2$ at $\gamma_R=0$ is illustrated in Figure~\ref{fig:B1}~(a). One can see that an increase of $\Delta_0$ leads to an increase of the lower, and to a decrease of the upper laser thresholds, together with a general decrease of the output power. Maximal output power $P_{max}$ and the pumping rate $w_m$ maximising the output power for different values of $\gamma_R$ are given in Figure~\ref{fig:B1}~(b) and (c) respectively. The parameters of the system were taken the same as in the Section~\ref{sec:YbMod}: $N=10^5$, $\gamma=2 \pi \times 43.5~{\rm mHz}$ , $\omega=2\pi \times 518.3~{\rm THz}$, $\g=50~{\rm s^{-1}}$, $\kappa=5\times 10^{5}~{\rm s^{-1}}$.

Also, it might be useful to derive the conditions for the existence of the non-zero field solution explicity. This solution exists, if the expression within the square brackets in equation (\ref{eq:B11}) is positive. If $w \gg \gamma, \gamma_R$, then this condition may be easily expressed as
\begin{equation}
\frac{\Gamma_c}{\Delta_0}>\sqrt{\frac{8}{\pi}}\frac{\exp(-\mathcal{B})}{{\rm erfc}(\sqrt\mathcal{B})}, \quad {\rm where} \quad \mathcal{B}=\frac{w^2}{8 \Delta_0^2},
\label{eq:B11a}
\end{equation}
and $\Gamma_c=N \g^2/\kappa$, as before. Note that at $\Delta_0 \rightarrow 0$, this condition transforms into $\Gamma_c>w$. Also, one may note that the right part of (\ref{eq:B11a}) can not exceed $\sqrt{8/\pi}$, which leads to the fundamental limit $\Gamma_c>\Delta_0 \sqrt{8/\pi}$.
\subsection{Stability of the non-zero steady-state solution}
\label{app:B2}

Here we build the function $\DD(\lambda)$ characterizing the stability of the steady-state solution (\ref{eq:B9}) -- (\ref{eq:B11}). Therefore, we have to construct the matrices $\GGG^{(j)}$, $\DDD^{(j)}$ and $(\lambda \III - \AAA^{(j)})^{-1}$. Matrix $\GGG^{(j)}$ can be easily found from (\ref{eq:19}), (\ref{eq:B5}) and (\ref{eq:B6}):
\begin{equation}
\GGG^{(j)}=\frac{i \g}{2}\left( 
\begin{array}{ccc}
-1 & 0 & 0 \\
0  & 1 & 0
\end{array}
\right). \label{eq:B12}
\end{equation}
Also, matrix $\DDD^{(j)}$ can be calculated from (\ref{eq:13}), (\ref{eq:14}), (\ref{eq:20}), (\ref{eq:B4}), (\ref{eq:B5}) and the steady-state solution (\ref{eq:B9}) -- (\ref{eq:B11}):
\begin{equation}
\DDD^{(j)}=\frac{i \g \langle \sig_z^j \rangle_{cw}}{2}\left( 
\begin{array}{cc}
1 & 0 \\
0  & -1\\
\displaystyle{\frac{i \g E}{\gamma_{\perp}-i \Delta_j}} & \displaystyle{\frac{i \g E}{\gamma_{\perp}+i \Delta_j}}
\end{array}
\right). \label{eq:B13}
\end{equation}
Here we took the arbitrary phase of the cavity field to be zero, which results in $\langle \a \rangle_{cw} = \langle \a^+ \rangle_{cw} = E$.

Calculation of the matrix  $(\lambda \III - \AAA^{(j)})^{-1}$ requires a little more effort:
\begin{widetext}
\begin{align}
(\lambda \III - \AAA^{(j)})^{-1}& = \frac{1}{(\lambda+\gamma_\parallel) [(\lambda + \gamma_\perp)^2+\Delta_j^2]+\g^2 E^2 (\lambda + \gamma_{\perp})} \times \nonumber \\
\nonumber \\
&\left[ 
\begin{array}{ccc}
\displaystyle{\frac{\g^2 E^2}{2}}+(\lambda + \gamma_{\parallel})(\lambda+\gamma_{\perp}-i \Delta_j) & 
\displaystyle{\frac{\g^2 E^2}{2}} &
\displaystyle{ \frac{i \g E}{2}} (\lambda + \gamma_\perp-i \Delta_j) \\
\\
\displaystyle{\frac{\g^2 E^2}{2}} & 
\displaystyle{\frac{\g^2 E^2}{2}}+(\lambda + \gamma_{\parallel})(\lambda+\gamma_{\perp}+i \Delta_j) &
- \displaystyle{\frac{i \g E}{2}} (\lambda + \gamma_\perp+i \Delta_j) \\
\\
i \g E (\lambda + \gamma_\perp- i \Delta_j) & - i \g E (\lambda + \gamma_\perp + i \Delta_j) &
(\lambda + \gamma_\perp)^2+\Delta_j^2
\end{array}
\right] \label{eq:B14}
\end{align}
\end{widetext}
Now we can calculate $\DD(\lambda)$ with the help of (\ref{e:47}). After some algebra, we can express the even part $\MMM(\Delta_j)$ of the matrix product $\GGG \cdot(\lambda \III - \AAA(\Delta_j))^{-1}\cdot \DDD(\Delta_j)$ in the form 
\begin{align}
\MMM(\Delta_j)&=\left(\begin{array}{cc} M_{d} (\Delta_j) & M_{nd} (\Delta_j) \\ M_{nd} (\Delta_j) & M_{d} (\Delta_j) \end{array} \right), \label{eq:B15}
\end{align}
where
\begin{align}
M_{d}(\Delta_j)&=\frac{\g^2d_0 (\lambda+\gamma_{\perp})}{4} \times \nonumber \\
&\left(\frac{\xi-\eta}{\zeta-\eta}\frac{1}{\Delta_j^2+\eta}+ \frac{\xi-\zeta}{\eta-\zeta}\frac{1}{\Delta_j^2+\zeta} \right), \label{eq:B16} 
\end{align}
\begin{align}
M_{nd}(\Delta_j)&=\frac{\g^4 E^2 d_0 \gamma_{\perp} (2 \gamma_\perp+\lambda)}{8(\lambda+\gamma_{\parallel}) (\zeta-\eta)} \times \nonumber \\
& \left(\frac{1}{\Delta_j^2+\zeta}-\frac{1}{\Delta_j^2+\eta} \right). \label{eq:B17}
\end{align}
Here we denoted 
\begin{align}
\xi&=\gamma_\perp^2 - \frac{E^2 \g^2 \lambda \gamma_{\perp}}{2 (\lambda+\gamma_{\perp})(\lambda+\gamma_{\parallel})}, \label{eq:B18} \\
\zeta&=(\lambda+\gamma_{\perp})^2+\frac{E^2 \g^2  (\lambda+\gamma_{\perp})}{\lambda+\gamma_\parallel}, \label{eq:B19} \\
\eta&=\gamma_{\perp}^2+\frac{E^2 \g^2 \gamma_\perp}{\Gamma_\parallel}. \label{eq:B20}
\end{align}

Now we should integrate the matrix $\MMM$ over the Gaussian distribution (\ref{e:33}). Introducing
\begin{equation}
\MM_{d,nd}=N\int_{-\infty}^{\infty} \frac{M_{d,nd}(\Delta_j)}{\sqrt{2 \pi}\Delta_0} \exp\left[-\frac{\Delta_j^2}{2\Delta_0^2} \right]d\Delta_j, \label{eq:B21}
\end{equation}
and using the function (\ref{e:43}), we can express the results of this integration in the form:
\begin{align}
\MM_d&=\frac{N \g^2d_0 (\lambda+\gamma_{\perp})}{8 \Delta_0^2 \sqrt{\pi} } 
\left(
\frac{\xi-\eta}{\zeta-\eta}\,  \Theta_\eta 
+ \frac{\xi-\zeta}{\eta-\zeta}\, \Theta_\zeta \right), \label{eq:B22} \\
\MM_{nd}&=\frac{N \g^4 E^2 d_0 \gamma_\perp (\lambda+2 \gamma_{\perp})}{16\, (\lambda+\gamma_\parallel)\, \Delta_0^2\, \sqrt{\pi} } \,
\frac{
\Theta_\zeta  - \Theta_\eta}{\zeta-\eta} , \label{eq:B23}
\end{align}
where we denoted $\Theta_{\zeta,\eta}=\Theta\left(\frac{\zeta,\eta}{2 \Delta_0^2}\right)$. Finally, with the help of (\ref{eq:B11}) one can show that
\begin{equation}
\Theta_\eta=\frac{2\Delta_0^2 \kappa \sqrt{\pi}}{\g^2 N \gamma_{\perp} d_0}.  \label{eq:B24}
\end{equation}

Therefore, we can express the function $\DD(\lambda)$ via the functions $\MM_d$ and $\MM_{nd}$ as
\begin{equation}
\DD(\lambda)_=\frac{(\lambda + \kappa/2 - \MM_d)^2-\MM_{nd}^2}{(\lambda+\kappa/2)^2}.  \label{eq:B25}
\end{equation}
Because the explicit form of $\DD(\lambda)$ is quite bulky, it is convenient to perform the stability analysis numerically, tracing the phase of $\DD(\lambda)$ along the contour, as described in Section~\ref{sec:lsa}.

\begin{figure*}[t]
\begin{center}
\resizebox{0.9\textwidth}{!}
{\includegraphics{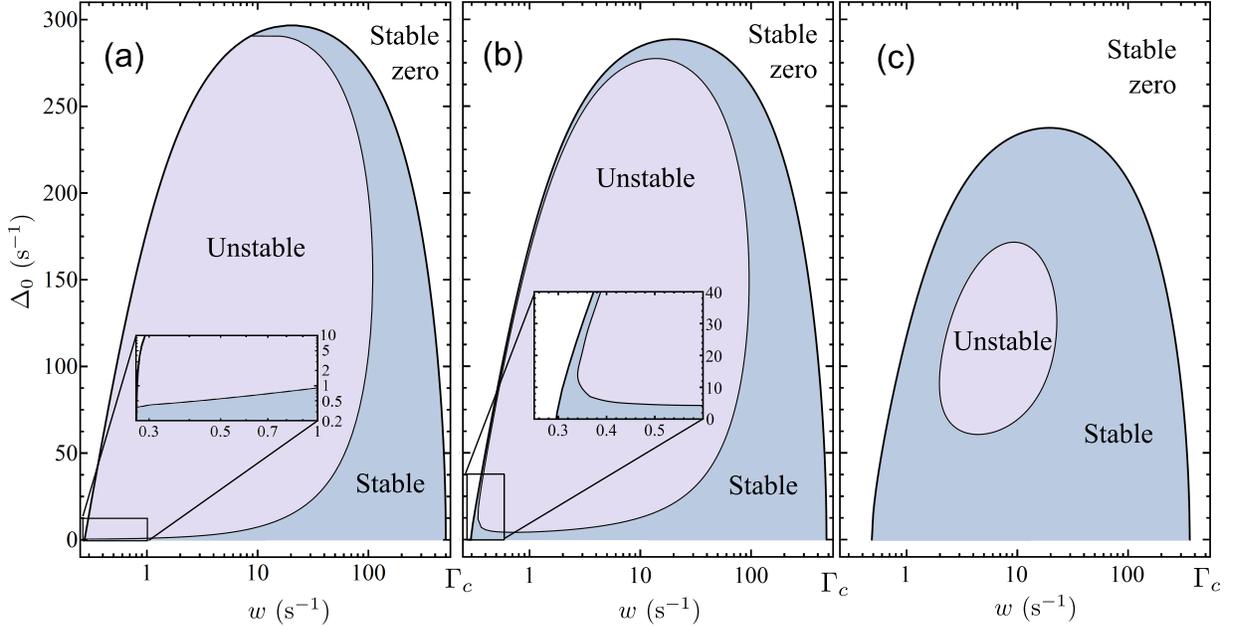}}
\end{center}
\caption{(color online) Domains of existence and stability of steady-state solutions for the two-level incoherently pumped laser with inhomogeneous broadening in the $(w,\Delta_0)$ plane for different values of $\gamma_R$ (a): $\gamma_R=0$; (b): $\gamma_R=10~s^{-1}$; (c): $\gamma_R=70~s^{-1}$. Insets are enlarged views of the stability domains at small $\Delta_0$ near the lower lasing threshold. Parameters of the cavity and the atomic ensemble are the same as for the Figure~\ref{fig:B1}.}
\label{fig:B2}
\end{figure*}

The existence and the stability domains for various steady-state solutions for the inhomogeneously broadened two-level atomic laser in the $(w, \Delta_0)$-plane are presented in Figure~\ref{fig:B2}. We should note that, although the presence non-zero inhomogeneous dephasing $\gamma_R$ suppresses the output power (see Figure~\ref{fig:B1}~(b)), it enlarges the stability domain. Particularly, at $\gamma_R=10~{\rm s^{-1}}$, the laser radiation becomes stable slightly above the lower laser threshold, but looses its stability with some increase of $w$, as it is shown in Figure~\ref{fig:B2}~(b). This result is in correspondence with \cite{Haken, Abraham85, Zhang85, Mandel85}. Further increase of $w$ will again stabilize the lasing, because the contribution of $w$ to the total decoherence rate becomes dominant. So, we have effectively two ``second laser thresholds'', the lower and the upper, lying between the lower and the upper first lasing thresholds. Note that an increase of $\gamma_R$ to higher values leads to a drastic reduction of the instability domain, as shown in Figure~\ref{fig:B2}~(c) for $\gamma_R=70~{\rm s^{-1}}$. At further increase of $\gamma_R$, the instability domain eventually disappears.
\subsection{Existence of non-zero field, and stability of the zero field solution}
Using the fact that $\Theta(y)$ is a strictly decreasing function of $y$ (see definition (\ref{e:43})), one can easily derive the condition for the existence of the non-zero field steady-state solution from (\ref{eq:B11}):

\begin{equation}
\gamma_{\perp} \Theta\left(\frac{\gamma_{\perp}^2}{2 \Delta_0^2}\right)> \frac{4 \Delta_0^2 \, \kappa \, \sqrt{\pi}}{\g^2 N  d_0}. \label{eq:B26}
\end{equation}

Let us investigate the stability of the zero-field solution (\ref{eq:B8}). As before, we build the function $\DD(\lambda)$. Matrices $\GGG^{(j)}$, $\DDD^{(j)}$ and $\AAA^{(j)}$ are
\begin{equation}
\GGG^{(j)}=\frac{i \g}{2}\left( 
\begin{array}{ccc}
-1 & 0 & 0 \\
0  & 1 & 0
\end{array}
\right), \label{eq:B27}
\end{equation}

\begin{equation}
\DDD^{(j)}=d_0 \frac{i \g}{2}\left( 
\begin{array}{cc}
1 & 0 \\
0  & -1 \\
0 & 0
\end{array}
\right), \label{eq:B28}
\end{equation}

\begin{equation}
\AAA^{(j)}=\left( 
\begin{array}{ccc}
-\gamma_{\perp}-i \Delta_j & 0 & 0 \\
0 & -\gamma_{\perp}+i \Delta_j  & 0 \\
0 & 0 & -\gamma_{\parallel}
\end{array}
\right). \label{eq:B29}
\end{equation}
According to (\ref{e:47}), we obtain
\begin{equation}
\DD(\lambda)=\left[1-\frac{N\, \g^2\, (\lambda+\gamma_\perp)\, d_0}{4\, \sqrt{\pi}\,  \Delta_0^2 \, (\kappa+2 \lambda)} \, \Theta\left(\frac{(\lambda+\gamma_{\perp})^2}{2 \Delta_0^2} \right) \right]^2. \label{eq:B30}
\end{equation}

Now we can show that the zero-field solution of the semiclassical equations is stable, if and only if the non-zero field solution does not exist. Indeed, if the inequality (\ref{eq:B26}) is fulfilled, then the equation 
\begin{equation}
(\lambda+\gamma_\perp) \Theta\left(\frac{(\lambda+\gamma_{\perp})^2}{2 \Delta_0^2}\right)=\frac{4\,  \Delta_0^2 \, (\kappa+2 \lambda)\, \sqrt{\pi}}{\g^2\, N\, d_0} 
\label{eq:B31}
\end{equation}
has a solution on the real positive semiaxis. This is a result of the fact that the right part of (\ref{eq:B31}) is a strictly increasing, whereas the left part is a strictly decreasing function of $\lambda$, approaching zero when $\lambda$ is approaching infinity. On the other hand, if (\ref{eq:B26}) is not fulfilled, there is no solution of (\ref{eq:B31}) with a positive real part of $\lambda$. To illustrate this, one can represent $(\lambda+\gamma_{\perp})/\sqrt{2 \Delta_0^2}=x+ i y$, and use the inequality
\begin{equation}
e^{(x+iy)^2}\mathrm{erfc}
(x+iy) \leq e^{x^2}{\rm erfc(x)} \quad \mathrm{for} \quad x>0 \label{eq:B32}
\end{equation}
which can easily be proven using the integral form of the complementary error function $\rm{erfc}$, see \cite{AbrStig}.


\end{document}